\documentclass[journal=jctc,manuscript=article]{achemso}
\usepackage{graphicx} % Required for inserting images
\usepackage{siunitx}
\usepackage{xcolor}
\usepackage{caption}
\usepackage{subcaption}
\usepackage{todonotes}
\usepackage{bigints}
\usepackage{algorithm}
\usepackage{algpseudocode}
\usepackage{hyperref}
\algdef{SE}[SUBALG]{Indent}{EndIndent}{}{\algorithmicend\ }%
\algtext*{Indent}
\algtext*{EndIndent}

\SectionNumbersOn

\usepackage{amsmath,amsfonts,amssymb}

\newcommand{\R}{\mathbb{R}}                                     % real numbers
\newcommand{\innerprod}[2]{\left\langle #1,\, #2 \right\rangle} % scalar product
\newcommand{\con}[2]{\cdot_{|#1,#2}}
\newcommand{\diff}{\mathrm{d}}                                   %'d' for integrals
\DeclareMathOperator*{\argmin}{arg\,min}

\newcommand{\bE}{\mathbb{E}}

\newcommand{\cK}{\mathcal{K}}
\newcommand{\cL}{\mathcal{L}}
\newcommand{\cP}{\mathcal{P}}

\newcommand{\BA}{\mathbf{A}}

\newcommand{\BG}{\mathbf{G}}

\newcommand{\BL}{\mathbf{L}}

\newcommand{\BR}{\mathbf{R}}

\newcommand{\BU}{\mathbf{U}}

\newcommand{\bh}{\mathbf{h}}

\title{Kinetically Consistent Coarse Graining using Kernel-based Extended Dynamic Mode Decomposition}
\author{Vahid Nateghi}
\affiliation{ 
Max-Planck-Institute for Dynamics of Complex Technical Systems, Magdeburg, Germany
}
\author{Feliks Nüske}
\affiliation{ 
Max-Planck-Institute for Dynamics of Complex Technical Systems, Magdeburg, Germany
}
\email{nateghi@mpi-magdeburg.mpg.de, nueske@mpi-magdeburg.mpg.de}

\begin{document}
\maketitle

\begin{abstract}
In this paper, we show how kernel-based models for the Koopman generator -- the gEDMD method -- can be used to identify coarse-grained dynamics on reduced variables, which retain the slowest transition timescales of the original dynamics. The centerpiece of this study is a learning method to identify an effective diffusion in coarse-grained space, which is similar in spirit to the force matching method. By leveraging the gEDMD model for the Koopman generator, the kinetic accuracy of the CG model can be evaluated. By combining this method with a suitable learning method for the effective free energy, such as force matching, a complete model for the effective dynamics can be inferred. Using a two-dimensional model system and molecular dynamics simulation data of alanine dipeptide and the Chignolin mini-protein, we demonstrate that the proposed method successfully and robustly recovers the essential kinetic and also thermodynamic properties of the full model. The parameters of the method can be determined using standard model validation techniques.
\end{abstract}

\section{Introduction}
Stochastic simulations of large-scale dynamical systems are widely used to model the behaviour of complex systems, with applications in computational physics, chemistry, materials science, and engineering. Many examples of such systems are high dimensional and subject to meta-stability, which means the system remains trapped in a set of geometrically similar configurations, while transitions to another such state are extremely rare. As a consequence, it becomes necessary to produce very long simulations in order to make statistically robust predictions. A prime example are atomistic molecular dynamics simulations (MD)~\cite{frenkel_understanding_2023} of macro-molecules, where meta-stability is typically caused by high energetic barriers separating deep potential energy minima~\cite{onuchic_theory_1997} . As a result, it requires specialized high-performance computing facilities to reach the required simulation times, or it may just not be feasible at all~\cite{karplus_molecular_1990} .

Coarse graining (CG) describes the process of replacing the original dynamical system by a surrogate model on a (much) lower-dimensional space of descriptors~\cite{das_low-dimensional_2006,clementi_coarse-grained_2008} , in such a way that certain properties of the original dynamics are preserved by the surrogate model. CG models can enable scientists to achieve much longer simulation times because of the reduced computational cost, while maintaining predictive capabilities of the full-order model. Setting up a CG model typically requires the following steps: first, the choice of a linear or non-linear mapping (CG map) from full state space to a lower-dimensional space, where the latter serves as the state space of the surrogate model. Second, definition of a parametric model class for the surrogate dynamics. Finally, determination of the parameters for that model class.

The first step is crucial to the CG model's success, and has been a very active area of research for a long time, see Refs.~\cite{rohrdanz_discovering_2013,wang_nonlinear_2018,sidky_machine_2020} for reviews on this topic. In this study, we only show examples of low-dimensional CG coordinates that have already been validated, while the problem of learning high-dimensional and fully transferrable CG models is left for future studies. Our focus is instead on the second and mainly the third step. CG models have often been parameterized using physically intuitive functional forms for the coarse-grained energy. More recently, much more general functional forms have been used for the CG parameters, which are then approximated by powerful model classes, such as deep neural networks or reproducing kernels~\cite{john_many-body_2017,zhang_deepcg_2018,wang_machine_2019} , which is the approach we follow in this paper. We study CG for reversible stochastic differential equations (SDE) with a Boltzmann-type invariant distribution, which is applicable to most popular simulation engines in MD, e.g. Langevin dynamics. Following the projection approach from Refs.~\cite{legoll_effective_2010,zhang_effective_2016} , we likewise parameterize the coarse-grained model as a reversible SDE, disregarding memory terms. 

The success of machine learning (ML) in recent years has led to the development of many powerful learning schemes for the parameters of a CG model. Examples are free energy learning~\cite{schneider_stochastic_2017} , and force matching~\cite{noid_multiscale_2008} , among others. Many of these learning methods are geared towards ensuring \emph{thermodynamic consistency}, which means that the surrogate model is trained to sample the marginalized Boltzmann distribution in CG space, thus ensuring accurate estimation of average quantities. Ensuring faithful reproduction of kinetic properties, such as time-correlation functions or transition timescales, is a much less developed topic, see Ref.~\cite{jin_bottom-up_2022} for an overview of prior work in this area. In this paper, we focus on the recovery of implied transition timescales~\cite{prinz_markov_2011} associated to meta stable states.

Implied transition timescales are defined in terms of the leading spectrum of the system's transfer or Koopman operator~\cite{davies_metastable_1982,dellnitz_approximation_1999,schutte_direct_1999,klus_data-driven_2018} . Equivalently, one may also consider the spectrum of the associated Koopman generator (Kolmogorov operator for SDEs) close to zero. This connection has been at the heart of the Markov state modeling (MSM) approach~\cite{prinz_markov_2011,sarich_approximation_2010,bowman_introduction_2014} and many important developments based on it~\cite{noe_variational_2013,mardt_vampnets_2018,wu_variational_2020} . The \emph{spectral matching} approach~\cite{nuske_coarse-graining_2019} was the first to make use of this connection, by first parameterizing the CG model as a linear expansion of fixed basis functions, and then solving a regression problem to recover the eigenvalues of the Koopman generator. This idea was formalized in Ref.~\cite{klus_data-driven_2020} , by suggesting to regress on a full matrix representation of the Koopman generator. The generator matrix can be estimated a priori by a data-driven algorithm called generator EDMD (gEDMD).

In this study, we significantly improve on the idea of leveraging the Koopman generator for the identification of coarse grained models. %generator-based spectral matching, by formulating the approach using kernel methods. Reproducing kernels~\cite{wendland_scattered_2004,christmann_support_2008} offer powerful model classes which have been applied in many areas of machine learning. Kernel-based approximation of the Koopman generator (kgEDMD) was first described in Ref.~\cite{klus_kernel-based_2020} We use the kgEDMD matrix and a (reduced and whitened) basis of kernel functions to formulate a learning problem for a generally state-dependent effective diffusion, which optimally approximates the Koopman generator. Our approach only requires simulation data of the full system, the choice of the kernel function, and measurements of the local diffusion, which is analogue to the local mean force in force matching. Combined with the latter, this approach provides complete access to the effective dynamics associated to a reversible system.
Our key contributions are the following:
\begin{itemize}
    \item We formulate a data-based learning problem for the effective diffusion of a coarse-grained SDE. This formulation is analogous to the force matching approach for the coarse-grained energy. Just as force matching relies on measurements of the local mean force, our approach rests on a similar quantity called local diffusion.
    \item We suggest to parametrize the diffusion by a basis of random Fourier features~\cite{rahimi_random_2007} , which form a widely-used approximation technique for reproducing kernels. Random features offer a compromise between representational power and computational efficiency. The only hyper-parameters to be tuned are those of the kernel function. The method is robust to statistical noise and ill-conditioning as it is based on a whitened and truncated basis set.
    \item We show how the gEDMD method can be leveraged to evaluate the kinetic consistency of the learned CG model by comparing its eigenvalues to those of the reference gEDMD matrix. This assessment does not require simulations of the CG model.
    % \item We show that diffusion learning and force matching can be combined into a single pipeline to completely parameterize a coarse-grained SDE for reversible systems.
    \item We show that kinetic and also theromodynamic consistency are achieved by the method using three test cases, a two-dimensional model system and molecular dynamics simulations of the alanine dipeptide and the Chignolin mini-protein. For the molecular systems, we consider CG models in low-dimensional reaction coordinate spaces, and we employ an overdamped assumption to simplify the learning process.
\end{itemize}

The structure of the paper is as follows: we introduce the required background on SDEs, coarse graining, and Koopman operator learning in Section~\ref{sec:theory}. Our learning framework is then presented in Section~\ref{sec:method}, while the numerical examples follow in Section~\ref{sec:examples}. An overview of the notation is provided in Table~\ref{tab:not}, supplementary information on simulation details and model selection is given in the Appendix.

\section{Theory}
\label{sec:theory}
In this section, we provide the necessary background on stochastic dynamics, data-driven modeling, and Koopman spectral theory. 

\subsection{Stochastic processes}
We consider a dynamical system described by a stochastic differential equation (SDE) 
\begin{equation}
dX_t = b(X_t) dt + \sigma(X_t) dW_t,
\label{eq:sde}
\end{equation}
where $b(X_t): \R^d\rightarrow \R^d$ is the drift vector field, $\mathbf{\sigma}(X_t): \R^d \rightarrow \R^{d\times d}$ is the diffusion field, and $W_t$ is a d-dimensional Brownian motion. We sometimes refer to the diffusion covariance matrix which is denoted as $a \in \mathbb{R}^{d\times d}$:
\begin{equation}
a(x) = \sigma(x) \sigma^\top(x).
\label{eq:diff-a}
\end{equation}
A standard example for Eq.~\eqref{eq:sde}, commonly used in molecular modeling, is overdamped Langevin dynamics
\begin{equation}
\label{eq:overdamped_langevin}
dX_t = -\dfrac{1}{\gamma}\nabla F (X_t) dt + \sqrt{2\beta^{-1}\gamma^{-1}} dB_t,
\end{equation}
where $F:\Omega \rightarrow \R$ is the potential energy, $\beta = (k_BT)^{-1}$ and $\gamma$ are constants corresponding to the inverse temperature and the friction, respectively. The invariant measure for $X_t$ in Equation~\eqref{eq:overdamped_langevin} is the Boltzmann distribution $\mu \propto \exp(-\beta F)$, and the dynamics are reversible with respect to $\mu$. More generally, a reversible SDE with invariant measure $\mu \propto \exp(-F)$ can be parameterized in terms of the generalized scalar potential $F: \R^d \mapsto \R$, and the diffusion covariance $a$, as follows~\cite{pavliotis_stochastic_2014}
\begin{equation}
dX_t = \left[-\dfrac{1}{2} a(X_t) \nabla F(X_t) + \dfrac{1}{2} \nabla \cdot a(X_t)\right] dt + \sigma(X_t) dW_t.
\label{eq:reversible_sde}
\end{equation}
We will only consider reversible SDEs in this paper, and make use of the parametrization in Equation~\eqref{eq:reversible_sde} when formulating learning methods.

\subsection{Koopman generator and spectral decomppsition}
Koopman theory~\cite{koopman_hamiltonian_1931,mezic_spectral_2005} lifts the dynamics in Equation~\eqref{eq:sde} into an infinite-dimensional space of observable functions to express the dynamics linearly. More precisely, the family of Koopman operators ${\cK^t}$ for stochastic dynamics is defined as 
\begin{equation}
\mathcal{K}^t \psi(x) = \mathbb{E}^{x}\left[\psi (X_t)\right] = \mathbb{E}\left[\psi(X_t)\,\vert\, X_0 = x \right],
    \label{eq:koop-sde}
\end{equation}
where $\bE\left[\cdot\right]$ denotes the expected value. The associated infinitesimal generator $\mathcal{L}$ is the time-derivative of the expectation value, which can be written as a linear differential operator:
% \begin{equation}
% \mathcal{L} \psi(x) = \frac{d}{dt}\bE^{x}\left[\psi (x_t)\right]|_{t=0}.
%     \label{eq:gen}
% \end{equation}
 
\begin{equation}
\label{eq:gen-sde}
\begin{split}
    \mathcal{L} \psi(x) &= b(x) \cdot \nabla\psi(x) + \frac{1}{2} a(x) :\nabla^2 \psi(x) \\
    &= \sum_{i=1}^d b_i(x) \frac{\partial}{\partial x_i}\psi(x) + \frac{1}{2}\sum_{i, j=1}^d a_{ij}(x) \frac{\partial^2}{\partial x_i \partial x_j} \psi(x),
\end{split}    
\end{equation}

where $a$ and $b$ are the diffusion and drift terms defined above, $\nabla^2[\cdot]$ is the Hessian matrix of a function, and the colon $:$ is a short-hand for the dot product between two matrices. For overdamped Langevin dynamics, Eq.~\eqref{eq:gen-sde} simplifies to
\begin{equation*}
    \mathcal{L} \psi(x) = -\frac{1}{\gamma}\nabla F(x) \cdot \nabla\psi(x) + \frac{1}{\gamma\beta} \Delta\psi(x).
\end{equation*}

The key quantity of interest are the eigenvalues and eigenfunctions of the generator. The study of spectral components of the generator helps us identify the long-time dynamics of the system. In molecular dynamics, we expect to find a number of eigenvalues close to zero, followed by a spectral gap. These low-lying eigenvalues are indicating the number of meta-stable states of the system, which are the macro states the system stays in the longest~\cite{davies_metastable_1982} . We write the eigenvalue problem for the generator as
\begin{equation}
    -\cL\psi_i = \lambda_i \psi_i.
   \label{eq:gen-eig-prob}
\end{equation}
The eigenvalues $\lambda_i$ of $-\cL$ must be non-negative, and the lowest eigenvalue $\lambda_1 = 0$ is non-degenerate~\cite{lelievre_partial_2016} : $0 = \lambda_1 < \lambda_2 \leq \lambda_3 \leq ...$ . We also refer to the eigenvalues as rates, and to their reciprocals as implied timescales~\cite{prinz_markov_2011}
\begin{equation}
    t_i = \frac{1}{\lambda_i}.
\end{equation}

\subsection{Coarse graining and projection}
One of the main motivations of this work is to learn an SDE representing the full dynamics~\eqref{eq:sde} on a coarse grained space. Coarse graining (CG) is realized by mapping the state space $\Omega$ onto a lower-dimensional space $\hat{\Omega} \subset \R^{d}$ by means of a smooth CG function $\xi$. We write $\nu~\propto \exp(-\beta F^\xi)$ for the marginal distribution of the full-space invariant measure $\mu$, where $F^\xi$ is the free energy in the CG space.

To define dynamics in the CG space, we use the \textit{conditional expectation operator}~\cite{legoll_effective_2010,zhang_effective_2016}
\begin{equation}
    \mathcal{P}\psi(z) = \frac{1}{\nu(z)}\bE^{\mu}[\psi(x)|\xi(x)=z],
\end{equation} 
where $z$ is a position in CG space. This operator calculates the average of a function $\psi$ over all $x \in \Omega$ whose projection onto CG space is the same point $z \in \hat{\Omega}$. Following the exposition in~\cite{zhang_effective_2016} , one can define the projected generator 
\begin{equation}
    \cL^{\xi} = \cP \cL \cP.
\end{equation} 
It turns out its action on a function $\varphi = \varphi(z)$ in CG space is given by
\begin{equation}
    \mathcal{L}^{\xi}(\phi)  = \mathcal{P}[\mathcal{L}\xi]\cdot \nabla_z \phi + \frac{1}{2}\mathcal{P}[\nabla \xi^T a \nabla \xi]:\nabla_z^2 \phi.
    \label{eq:sde_cg}
\end{equation}
As one can see, $\mathcal{L}^{\xi}$ is of the same form as the original generator $\mathcal{L}$ in Equation~\eqref{eq:gen-sde}, and indeed it is the generator of an SDE $Z_t$ on $\hat{\Omega}$
\begin{equation}
	dZ_t = b^{\xi}(Z_t) dt + \sigma^{\xi}(Z_t) dW_t.
    \label{eq:sde-cg}
\end{equation}
The effective drift and diffusion coefficients are given in analytical form by
\begin{equation}
\label{eq:coeffs_cg_sde}
	b^{\xi}(z) = \mathcal{P}\left(\mathcal{L}\xi\right)(z)  \hspace{1cm} a^{\xi}(z) = \mathcal{P}\left(\nabla \xi^T a \nabla \xi\right)(z),
\end{equation}
and the practical task of coarse graining is to approximate them numerically.

\subsection{Generator EDMD}
 Numerical approximations to the infinitesimal generator $\cL$ can be obtained by a data-driven learning method called generator extended dynamic mode decomposition ~\cite{klus_data-driven_2020}(gEDMD). Given a finite set of scalar basis functions $\psi(x) = \{\psi_1(x), ..., \psi_n(x)\}$, and training data $\{x_l\}_{l=1}^m$ sampled from the invariant measure $\mu$, we form the matrices
 \begin{align*}
     \mathbf{\Psi} &= \left[\psi_i(x_l) \right]_{i,l}, & \cL\mathbf{\Psi} &= \left[\cL\psi_i(x_l) \right]_{i,l},
 \end{align*}
 using the analytical formula~\eqref{eq:gen-sde} to evaluate the second of these matrices. The solution of a linear regression problem leads to the matrix approximation
 \begin{equation}
 \label{eq:gen-ag}
     \mathbf{L} = \hat{\BG}^{-1}\hat{\BA},
 \end{equation}
 where
 \begin{align}
    \hat{\BA}_{ij} &= \frac{1}{m}\sum_{k=1}^m \psi_i(x_k) \cL\psi_j(x_k), &
    \hat{\BG}_{ij} &= \frac{1}{m}\sum_{k=1}^m \psi_i(x_k)\psi_j(x_k).
    \label{eq:gen-AG-est}
\end{align}
These matrices are empirical estimators of the following mass, stiffness, and generator matrices
\begin{align}
\BA_{ij} &= \innerprod{\psi_i}{\cL\psi_j}_\mu & \BG_{ij} &= \innerprod{\psi_i}{\psi_j}_\mu, & \mathbf{L} &= \mathbf{G}^{-1}\mathbf{A}.
    \label{eq:gen-approx}
\end{align}
The empirical mass matrix $\hat{\mathbf{G}}$ is often ill-conditioned. A standard approach to circumvent this is to perform a whitening transformation based on removing small eigenvalues
\begin{align}
    \widehat{\BG} &= \BU \Sigma \BU^\top, &
    \BR &= \BU\Sigma^{-0.5} \in  \mathbb{R}^{n\times r}, &
    \hat{\BL}_r &= \BR^\top \hat{\BA}\hspace{0.1cm} \BR,
    \label{eq:gen-red}
\end{align}
in which $r\leq n$. Here, $\BR$ is a transformation matrix mapping the original basis to the reduced basis 
\begin{equation}
\label{eq:reduced_basis}
    \bh(x) = \BR^\top \psi(x).
\end{equation}
Dominant eigenvalues of the generator can be computed by diagonalizing the matrix $\hat{\BL}$ or $\hat{\BL}_r$.

For arbitrary stochastic dynamics, the computation of $\BA$ involves a second-order differentiation as shown in Equation~\eqref{eq:gen-sde}. However, if the stochastic dynamics are reversible, only first-order derivatives are required to compute the matrix $\BA$, as the generator satisfies the following integration-by-parts formula
\begin{equation}
    \BA_{ij} = \left< \psi_i, \mathcal{L}\psi_j\right>_{\mu} = -\frac{1}{2} \int \nabla \psi_i \sigma \sigma^\top \nabla \psi^\top_j\, \diff \mu.
    \label{eq:gen-A-rev}
\end{equation}
Importantly, if the basis functions are actually defined in a CG space $\hat{\Omega}$, that is $\psi_i(x) = \psi_i(\xi(x))$, then by the chain rule the matrix $\BA$ can be written as
\begin{equation}
\begin{split}
    \BA_{ij} &= -\frac{1}{2} \int \nabla_x \psi_i \sigma \sigma^\top \nabla_x \psi^\top_j\, \diff \mu = - \frac{1}{2} \int (\nabla_z \psi_i \nabla_x \xi) \sigma \sigma^\top (\nabla_x \xi^\top\nabla_z \psi^\top_j)\, \diff \mu  \\ &= - \frac{1}{2} \int (\nabla_z \psi_i) (\nabla_x \xi \sigma \sigma^\top \nabla_x \xi^\top)(\nabla_z \psi^\top_j)\, \diff \mu.
\end{split}
    \label{eq:gen-A-rev-exp}
\end{equation}
We refer to the matrix 
\begin{equation}
\label{eq:local_diffusion}
    a^\xi_{\rm loc}(x) = \nabla_x \xi \sigma \sigma^\top \nabla_x \xi^\top \in \R^{d \times d}
\end{equation} 
as \textit{local diffusion}, and note that it is independent of the basis functions. It can therefore be computed a priori in numerical calculations.

\subsection{Random Fourier features}
The gEDMD algorithm requires choosing a set of basis functions $\psi(x)$. In this work, we use random Fourier features (RFFs), which are defined as
\begin{equation}
\label{eq:random_features}
    \psi(x) = \{\cos(\omega_1^\top x),\,\sin(\omega_1^\top x), ..., \cos(\omega_n^\top x),\,\sin(\omega_n^\top x)\}.
\end{equation}
The vectors $\omega_1, \ldots, \omega_n$ are random frequency vectors drawn from a spectral distribution $\rho$. RFFs provide a low-rank approximation to a reproducing kernel function~\cite{rahimi_random_2007} , and can therefore generate a powerful basis without the need for manual basis set design. The precise relation between kernel-based gEDMD and random features was presented in Ref.~\cite{nuske_efficient_2023}. In the following applications, we use the spectral measure associated to a Gaussian squared exponential kernel with bandwidth parameter $\gamma$
\begin{equation}
      k(x_i,x_j)  = \exp\left(-\frac{\|x_i-x_j\|)^2}{2\gamma^2}\right),
      \label{eq:ker-gauss}
\end{equation}
or to a periodic Gaussian kernel~\cite{duvenaud_automatic_2014} on periodic domains, such as dihedral coordinates. 

\section{Methods}
\label{sec:method}

The main proposal in this work is a method to identify a diffusion field for the coarse-grained dynamics based on data-driven approximations to the generator. We also show how the quality of the diffusion field in terms of reproducing dominant eigenvalues can be assessed, and finally, how the diffusion can be complemented to identify the complete CG dynamics. We recall that the dynamical equation in CG space is given by~\eqref{eq:sde-cg}, where the drift can be written as follows because of reversibility~\cite{pavliotis_stochastic_2014}
\begin{equation}
	b^{\xi} = -\dfrac{1}{2} a^{\xi} \nabla_z F^{\xi} + \dfrac{1}{2} \nabla \cdot a^{\xi}.
	\label{eq:drift_app}
\end{equation}  

\subsection{Diffusion Learning}
By Equation~\eqref{eq:coeffs_cg_sde}, the analytical effective diffusion $a^\xi$ is the best-approximation of the local diffusion $a^\xi_{\mathrm{loc}}$ by a (matrix-valued) function on the CG space. Hence, we can solve the following data-based minimization problem
\begin{equation}
\label{eq:minimization_diff}
    a^{\xi} = \underset{a = a(z)}{\argmin} \frac{1}{m}\sum^m_{i=1} \left\|a(\xi(x_i)) - a^{\xi}_{\mathrm{loc}}(x_i) \right\|_F ^ 2,
\end{equation}
where $\|\cdot \|_F$ is the Frobenius norm for matrices. We parametrize the diffusion field $a^\xi$ element-wise as a linear combination of the reduced RFF basis
\begin{equation}
\label{eq:param_effective_diff}
    (a^\xi_\alpha)_{ij}(z) = \sum_{l=1}^r \alpha^{ij}_l h_l(z) = \alpha \con{3}{1} \bh(z) = \alpha \con{3}{1} \BR^\top \psi(z),
\end{equation}
where we view the coefficient array $\alpha$ as a third-order tensor of dimension $d \times d \times r$, and the symbol $\con{i}{j}$ denotes contraction over indices $i$ and $j$ of two arrays. The parametrization must be symmetric, i.e. $a^\xi_{ij} = a^\xi_{ji}$, and we may also choose to set specific elements to zero, for example to enforce a diagonal diffusion field. With the parametrization~\eqref{eq:param_effective_diff}, the minimization problem~\eqref{eq:minimization_diff} becomes a linear regression problem that can be directly solved, potentially after regularization.

\subsection{Recovery of Spectral Properties}
After solving the minimzation problem~\eqref{eq:minimization_diff}, we can make use of the gEDMD method to assess the dynamical properties of the learned SDE in CG space. Using the integration-by-parts formula~\eqref{eq:gen-A-rev}, the elements of the reduced generator matrix corresponding to the diffusion field~\eqref{eq:param_effective_diff} with coefficient array $\alpha$ are
\begin{equation*}
    \innerprod{h_r}{\cL^\alpha h_s}_\nu = -\frac{1}{2}\int \nabla h_r(z) a^\xi_\alpha(z) \nabla h_s(z)^\top.
\end{equation*}
In matrix notation, this leads to the following explicit formula for the parametrized generator matrix, which can be computed directly without resorting to numerical simulations of the CG dynamics
\begin{equation}
\label{eq:red_gen_alpha}
    \hat{\BL}_r^\alpha = \sum_{i=1}^m \BR^\top \nabla_z \psi(z_i) \left[\alpha \con{3}{1} \BR^\top \psi(z_i)\right] \nabla_z \psi(z_i)^\top \BR.
\end{equation}
Properties inferred from the matrix $\hat{\BL}^\alpha_r$ can be compared to those obtained from the original gEDMD matrix $\hat{\BL}_r$ estimated off the full-space simulation data. For example, diagonalization of both $\hat{\BL}_r$ and $\hat{\BL}^\alpha_r$ leads to estimates $\lambda_i$ and $\lambda_i^\alpha$ for the dominant generator eigenvalues, which can be systematically compared. We mainly resort to comparing dominant eigenvalues in the examples below, but we point out that a more detailed assessment is possible: for instance, by computing matrix exponentials $\exp\left(t \hat{\BL}_r\right)$ and $\exp\left(t \hat{\BL}^\alpha_r\right)$, time-correlation functions can also be evaluated.

\subsection{Learning the Effective Potential}
We have seen that the accuracy of the effective diffusion field largely determines the dynamical properties of the coarse grained dynamics. In order to run simulations of the CG dynamics, and to ensure thermodynamic consistency, the effective potential $F^\xi$ also must be learned in a parametric form. This is not the main focus of our study, hence we just point out a few options.
A well-known and generally applicable technique is \textit{force matching}~\cite{noid_multiscale_2008} , which is based on the following minimization problem for the effective force
\begin{equation}
    \nabla_z F^{\xi} = \underset{g = g(z)}{\argmin} \frac{1}{m}\sum^m_{i=1} \left\|g(\xi(x_i)) - f^{\xi}_{\text{lmf}}(x_i) \right\| ^ 2,
    \label{eq:minimization-fm}
\end{equation}
where $f^{\xi}_{\text{lmf}}$ is called \textit{local mean force} and defined as follows
\begin{align}
	f^{\xi}_{\text{lmf}} &=  -\nabla_x F \cdot G^{\xi} + \nabla_x \cdot G^ {\xi}, &
	G^{\xi} &= \nabla_x \xi [(\nabla_x \xi)^\top\nabla_x \xi] ^{-1}.
 \label{eq:f_lmf}
\end{align} 
We point out the similarity to~\eqref{eq:minimization_diff}, which also led us to the name local diffusion for $a^\xi_{\mathrm{loc}}$. The effective potential can be parametrized as a linear combination of basis functions, such as random features, or as a deep neural network~\cite{wang_machine_2019} . In low-dimensional CG spaces, it also possible to approximate the projected invariant distribution $\nu$ as a linear combination of kernel functions centered at the data sites, known as \emph{kernel density estimate} (KDE)~\cite{parzen_estimation_1962} . Since we only consider low-dimensional CG spaces here, we opt for the KDE option in the examples below.

\begin{algorithm}[H]
\caption{Learning Effective Dynamics}\label{alg:cap}
\setlength{\tabcolsep}{.5ex}
  \begin{tabular}{ll}
    \textbf{Input:} & full space data $\{x_k\}_{k=1}^m$ in $\R^d$, CG map $\xi$, kernel function with spectral measure $\rho$, \\
    & Truncation rule for $r$ in Equation~\eqref{eq:gen-red}, regularization parameters
    % \textbf{Output:} & TT approximation of the transformed data tensor $\PSI(X)$
  \end{tabular}
  \hrule\vspace{0.2cm}
\begin{algorithmic}[1]
    \State Diffusion learning 
    \Indent
    \State Compute local diffusion $a^\xi_{\rm loc}$ as in Equation~\eqref{eq:local_diffusion}.
    \State Generate random feature basis $\psi$ as in Equation~\eqref{eq:random_features}.
    \State Compute reduced basis $\bh$ according to~\eqref{eq:reduced_basis}.
    \State Compute reduced generator matrix $\hat{\BL}_r$ as in Equation~\eqref{eq:gen-red}.
    \State Perform the minimization problem as in Equation~\eqref{eq:minimization_diff}.
    \State Compute learned generator matrix $\hat{\BL}_r^\alpha$ by~\eqref{eq:red_gen_alpha} and compare its properties to $\hat{\BL}_r$.
    \EndIndent
    \State Learning the full CG dynamics
    \Indent
    \State Learn effective potential $F^\xi$ by KDE or force matching~\eqref{eq:minimization-fm}.
    \State Effective drift is given by~\eqref{eq:drift_app}.
    \EndIndent
\end{algorithmic}
\end{algorithm}

\subsection{Overdamped Models for Molecular Systems}
For the molecular examples considered in Sections~\ref{subsec:alanine_dipeptide} and~\ref{subsec:chignolin}, the dynamics are not governed by the reversible overdamped dynamics~\eqref{eq:overdamped_langevin}. Nevertheless, it can be expected that their position space dynamics are similar to the overdampled process after a re-scaling of time. This can be rigorously shown for the underdamped dynamics used in Section~\ref{subsec:alanine_dipeptide}~\cite{lelievre_free_2010} . Fitting an overdamped model is simpler than modeling the full non-reversible dynamics if a non-linear CG map is used. Therefore, we compute the local diffusion as
\begin{equation}
    a^\xi_{\rm loc}(x) = \nabla_x \xi(x) \frac{2}{\beta \gamma }M^{-1} \nabla_x \xi^\top(x),
    \label{eq:loc-diff}
\end{equation}
where $M$ is the diagonal mass matrix of all atoms, $\beta$ is the inverse temperature, $\gamma$ is the friction, and $\nabla_x \xi(x)$ is the Jacobian of the CG map. This way of computing the local diffusion corresponds to a reversible overdamped Langevin process in full position space, meaning that the resulting CG model is an approximation to that same overdamped process. This also means that the eigenvalues and timescales of the overdamped model may differ from the ones obtained by directly approximating the full non-reversible process, but the meta-stable states will remain the same.

\section{Examples}
\label{sec:examples}
To show the effectiveness of the proposed method, we apply it to a two-dimensional model system defined by the Lemon-slice potential, and to MD simulation data of the alanine dipeptide and of the mini-protein Chignolin, which are widely-used test cases in molecular dynamics.\\
The optimization problems for learning the effective force and diffusion are solved by standard minimizers from \textit{scipy}'s optimization toolbox. Note that in case the sought after object in the optimization blocks are decoupled, e.g. under the assumption that the effective diffusion is diagonal, we replace the minimizers with a simple least-square problem, making the step much cheaper.   

\subsection{Lemon-Slice potential}
\subsubsection{System introduction} 
The Lemon-slice system is governed by overdamped Langevin dynamics in Equation~\eqref{eq:overdamped_langevin} with the following potential $F$
\begin{equation}
	F(x, y) = F(r,\phi) = \cos(4\phi) + 10(r-1)^2,
\end{equation}
where $r$ and $\phi$ are polar coordinates. The energy landscape of the system is shown in Figure~\ref{fig:lemon-system}a. To form the SDE for this example, we consider a diagonal state-dependent diffusion field $\sigma(x)$ defined as 
\begin{equation}
        \sigma(x) = 
         \begin{bmatrix}
            \sqrt{\frac{2}{\beta}(\sin(\phi)+1.5)}&  0\\
            0& \sqrt{\frac{2}{\beta}(\sin(\phi)+1.5)} \\
        \end{bmatrix} 
    \end{equation}
where $\beta=1$ is the inverse temperature. Using the Euler-Maruyama scheme at discrete integration time step $dt=10^{-3}$ for integration of the SDE, we collect the training data for learning. For the sake of validation and showing the robustness of the method, we produce $5$ independent experiments, each with length of $m=10^5$ time steps. We further down sample them to $1000$ samples each for learning effective force and diffusion. \\

As shown in previous studies, the polar angle $\phi$ is a suitable CG coordinate for this system, as it resolves all four meta-stable states
\begin{equation}
	\xi(x, y) = \phi.
\end{equation}
For this system, analytical expressions for the effective drift and diffusion along $\xi$ can be obtained by a slight modification of the results in Ref.~\cite{nuske_spectral_2021} , and serve as reference values.

We apply our learning method with random Fourier features on the reaction coordinate $\xi$, to identify the generator eigenvalues and meta-stable states and, subsequently, to identify an effective dynamics along $\xi$ using Algorithm~\ref{alg:cap}. As the polar angle is a periodic reaction coordinate (RC), we use the spectral measures associated to both a periodic and non-period Gaussian kernel and compare them.
The kernel bandwidth in either versions of Gaussian kernel is optimized using cross validation based on the VAMP-score \cite{wu_variational_2020} . Details on the VAMP-score analysis are reported in the appendix.

\subsubsection{Meta-stability analysis}
\begin{figure}
    \begin{center} 
    \begin{subfigure}[c]{0.45\textwidth}  
    \includegraphics[width=\textwidth]{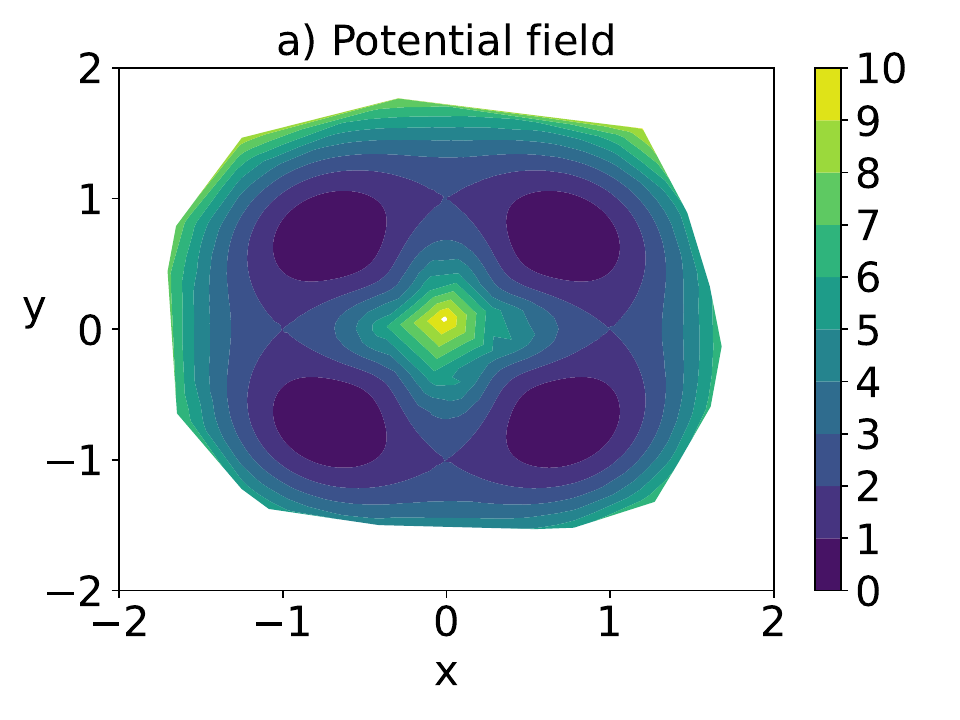}
    % \caption{}
    \label{fig:lemon-pot}
    \end{subfigure}
    \begin{subfigure}[c]{0.45\textwidth}  
    \includegraphics[width=\textwidth]{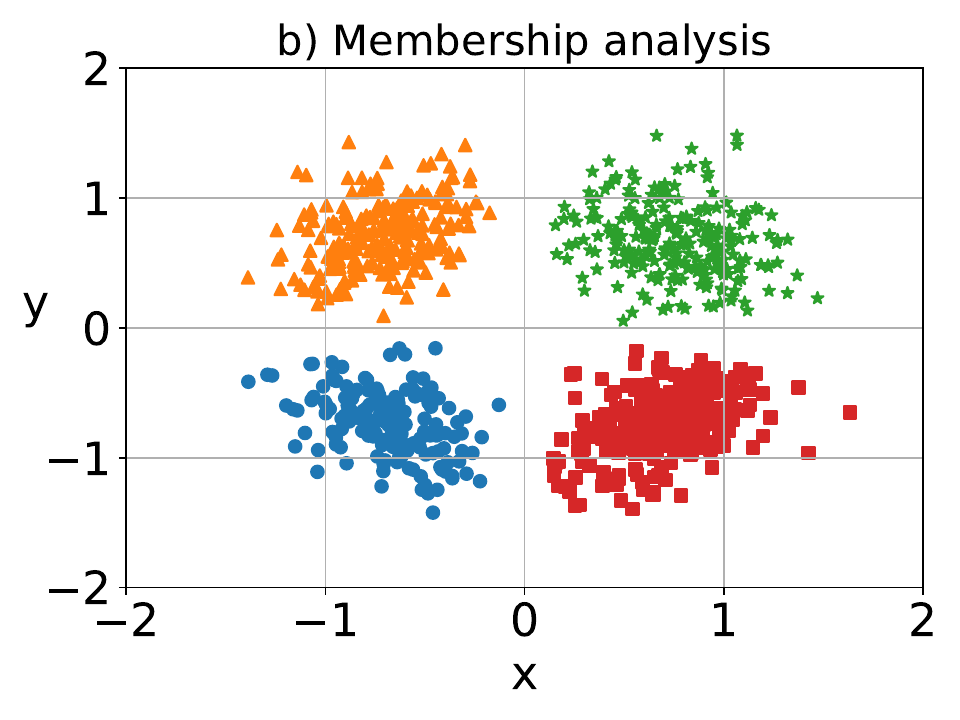}
    % \caption{}
    \label{fig:lemon-member}
    \end{subfigure}
    \begin{subfigure}[c]{0.45\textwidth}		\includegraphics[width=\textwidth]{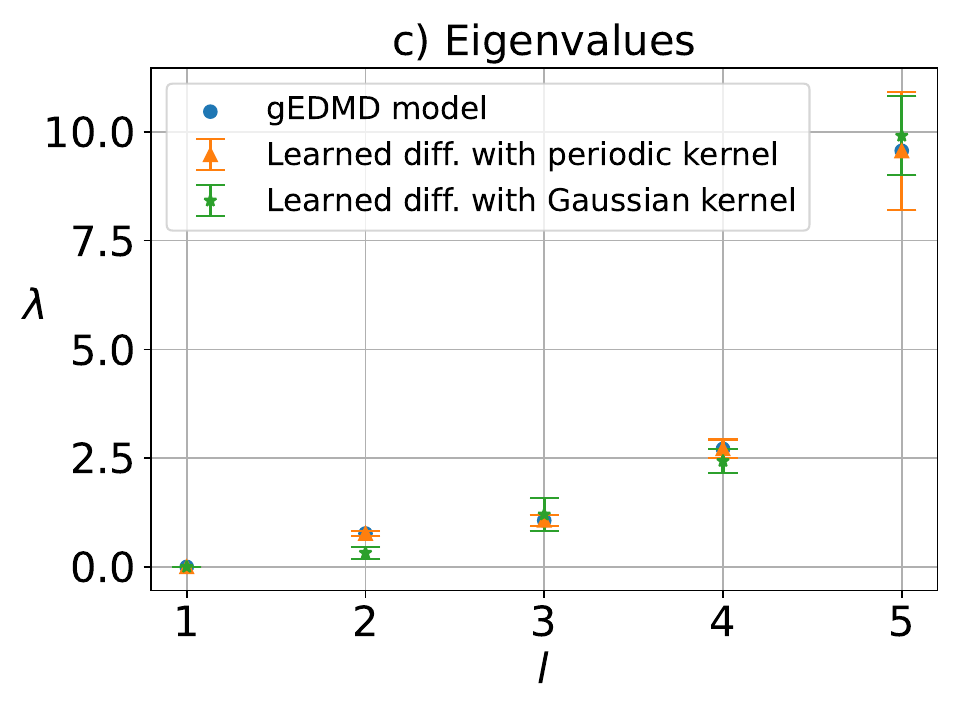} 
    % \caption{}
    \label{fig:lemon-eigs-gauss1}
    \end{subfigure}
    \begin{subfigure}[c]{0.45\textwidth}
        \includegraphics[width=\textwidth]{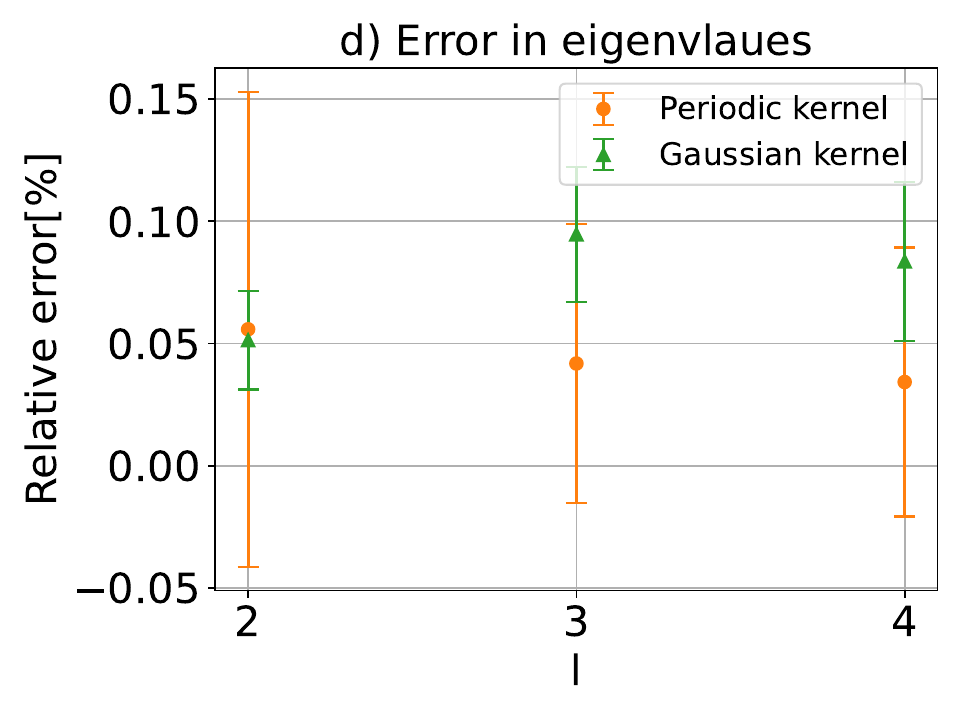}
        % \caption{}
        \label{fig:lemon-eigs-peri2}
    \end{subfigure} 
    \caption{Approximation of generator for the Lemon slice system. Potential field in (a). Membership analysis in (b) using $1000$ samples. The dominant eigenvalues of the reference generator $\hat{\BL}_r$ and the learned generator $\hat{\BL}_\alpha^\xi$ built upon the learned effective diffusion, using Gaussian and periodic Gaussian kernels, in (c). The relative error of these eigenvalues compared to the reference is shown in (d).}
    \label{fig:lemon-system}
    \end{center}
\end{figure}

Figure~\ref{fig:lemon-system}c shows the leading eigenvalues obtained from the generator matrix $\hat{\BL}_r$. As one notices, there are four dominant eigenvalues followed by a gap. These four eigenvalues are corresponding to the four minima in the potential field. Having determined the eigenvectors of the generator, we can perform robust Perron Cluster Cluster Analysis (PCCA+)~\cite{deuflhard_robust_2005} algorithm to assign to each sample point its membership to each meta-stable state. Figure~\ref{fig:lemon-system}b shows that the four potential minima are perfectly recovered in this way.
A comparison of the leading eigenvalues of the reference model $\hat{\BL}_r$ and the learned matrix $\hat{\BL}_\alpha^\xi$ for the optimal parameters $\alpha$ is shown in Figure~\ref{fig:lemon-system}c. Both choices of the kernel function lead to satisfactory results, the periodic kernel provides slightly higher accuracy in approximation of the generator eigenvalues. Note that the kernel bandwidth is tuned for each kernel function separately. 

\subsubsection{Analysis of the CG dynamics}
\begin{figure}
    \begin{center}
    \begin{subfigure}[c]{0.45\textwidth}
        \includegraphics[width=\textwidth]{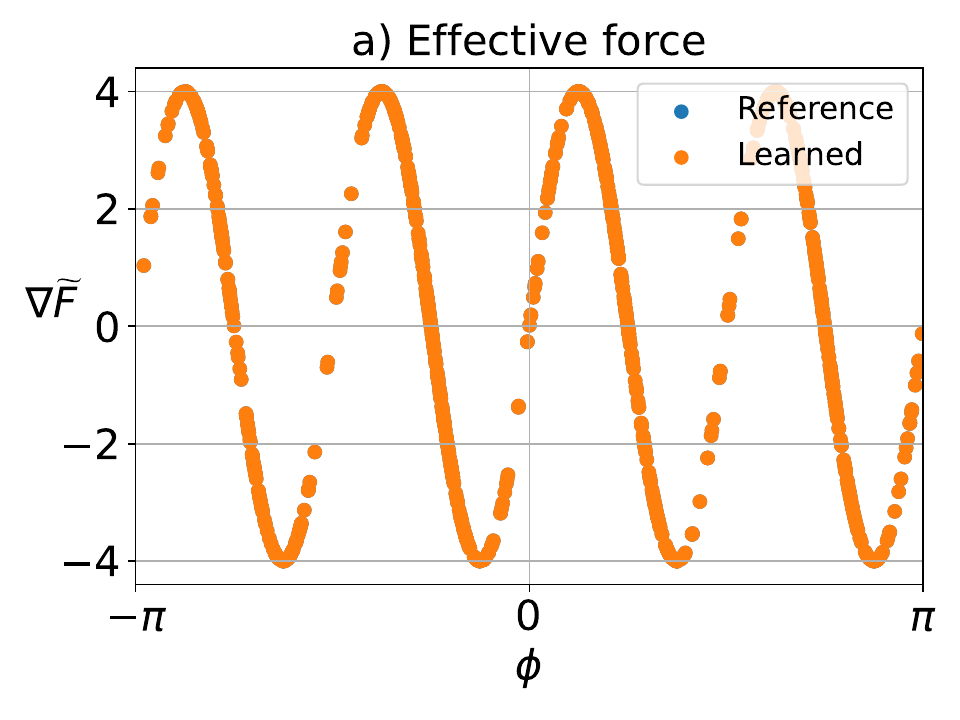}
        % \caption{}
        % \label{fig:lemon-learned-force}
    \end{subfigure}
    \begin{subfigure}[c]{0.45\textwidth}
        \includegraphics[width=\textwidth]{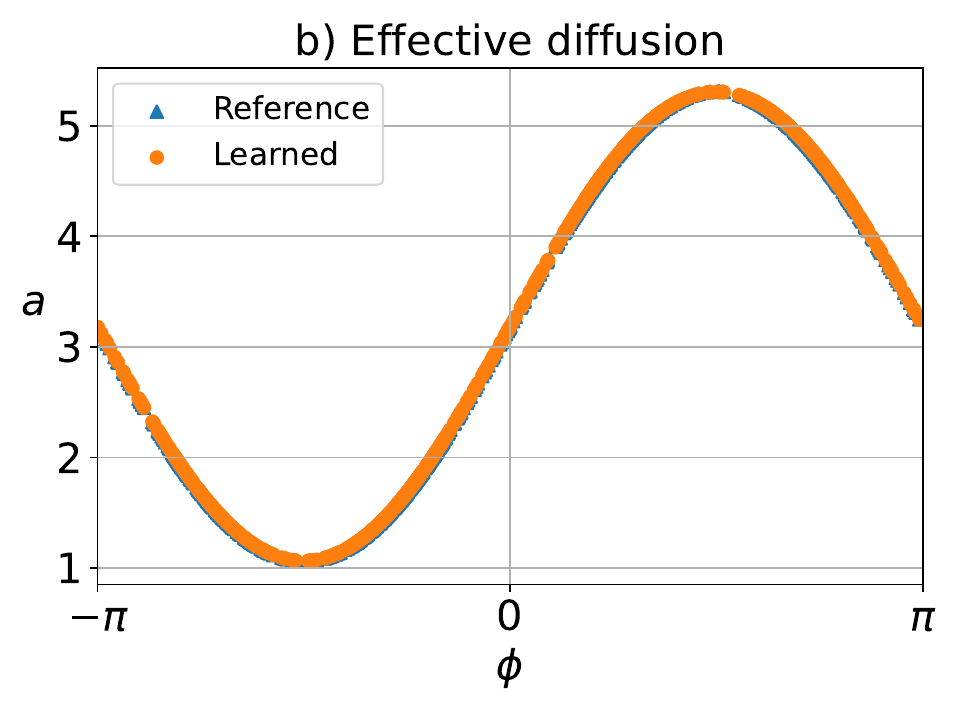}
        % \caption{}
        % \label{fig:lemon-learned-diff}
    \end{subfigure} \\
    \begin{subfigure}[c]{0.45\textwidth}
        \includegraphics[width=\textwidth]{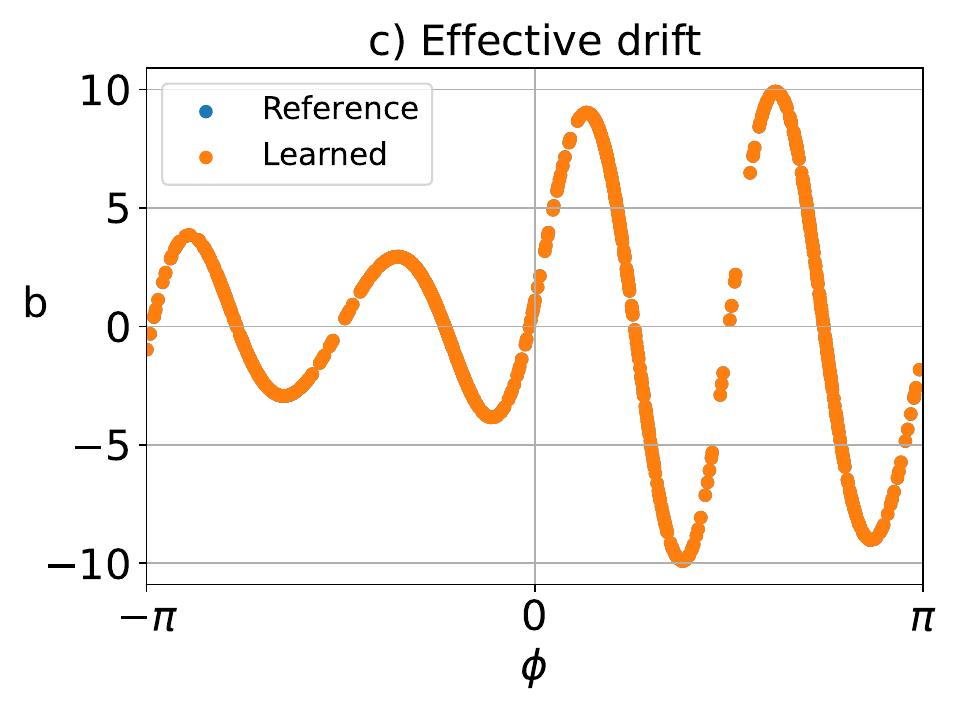}
        % \caption{}
        % \label{fig:lemon-learned-drift}
    \end{subfigure} 
    \begin{subfigure}[c]{0.45\textwidth} 
        \includegraphics[width=\textwidth]{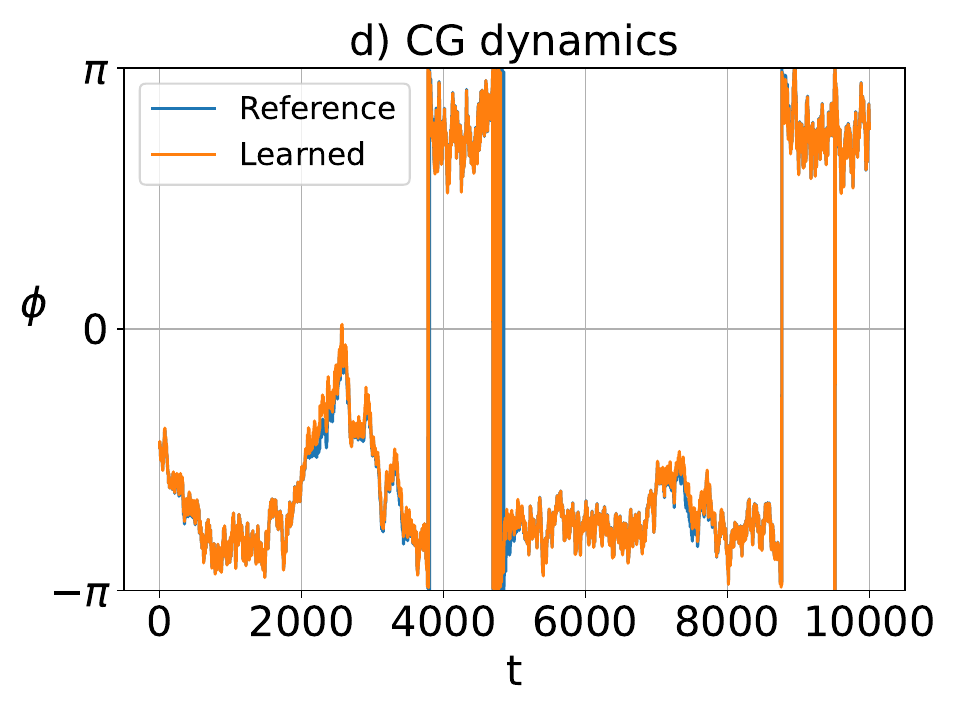}
        % \caption{}
    \end{subfigure} 
    \caption{Application of Algorithm~\ref{alg:cap} to identify angular dynamics for the Lemon-slice system. Effective force in (a), effective diffusion in (b), effective drift in (c), and integration of an example trajectory, using both the reference and learned SDE in (d).}
    \label{fig:lemon-learned-sde}
    \end{center}
\end{figure}

The learned generator providing the eigenvalues reported above is built upon the effective diffusion shown in Figure~\ref{fig:lemon-learned-sde}b, which is almost perfectly following the reference. Furthermore, we perform the force matching as well and obtain the effective force in the CG space shown in Figure~\ref{fig:lemon-learned-sde}a. From the effective force and diffusion, the effective drift can be obtained according to Equation~\eqref{eq:drift_app}, which is also compared against the analytical expression in Figure~\ref{fig:lemon-learned-sde}c, likewise showing very good agreement.\\

With the effective drift and diffusion fields, we are able to simulate the learned SDE governing the CG coordinate. We use the Euler-Maruyama scheme to integrate the learned and reference SDEs with integration time step of $dt=10^{-3}$. Figure~\ref{fig:lemon-learned-sde}d shows two trajectories of the CG coordinate $\phi$ for both dynamics for $10^4$ time steps, using the same Brownian motion for both trajectories. The propagated learned system follows the reference closely, with both systems staying long times in each meta-stable state, and rarely swapping in between those. Combined, the results above demonstrate that the proposed method can approximate the full system's meta-stable sets well, and identify a suitable SDE for CG dynamics which is accurate even on the level of individual trajectories.\\

\begin{figure}
    \begin{center} 
    \includegraphics[width=0.5\textwidth]{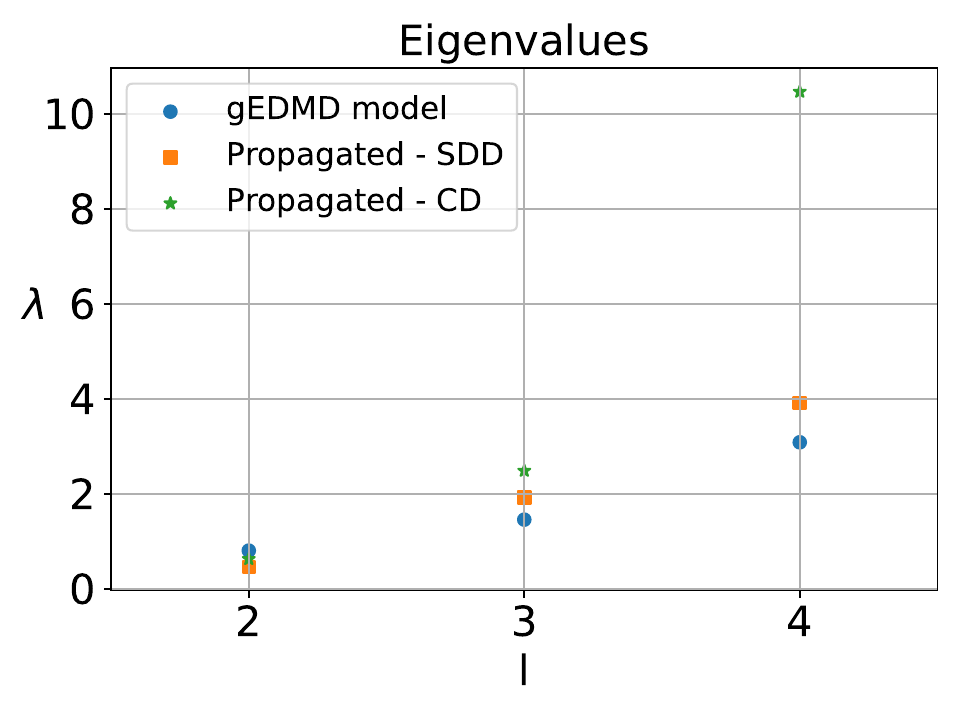}
    \caption{Dominant eigenvalues of the generator, using models built on simulation data of the learned coarse-grained dynamics with state-dependent diffusion (SDD, orange) and with constant diffusion (CD, green). As a comparison, we show the eigenvalues of the generator $\hat{\BL}_r$ using the original dataset (blue). Note that the first eigenvalue is omitted as it is zero.}
    \label{fig:lemon-prop-eigs}
    \end{center}
\end{figure}

As a final analysis, we compare the properties of the learned CG model with variable diffusion to those of a CG dynamics with constant diffusion, in order to demonstrate the necessity of allowing a state-dependent diffusion. We set the effective diffusion for the constant model to $a = \frac{2}{\beta} = 2$. We propagate the corresponding SDEs for a sufficiently large span of time, and estimate a new generator EDMD model based on these simulations. Figure \ref{fig:lemon-prop-eigs}, shows the eigenvalues of the generator for these cases compared to the learned generator built upon the original dataset. The result shows that learning a state-dependent diffusion is necessary to recover the original system's leading eigenvalues.

\subsection{Alanine dipeptide}
\label{subsec:alanine_dipeptide}
\subsubsection{System introduction}
\begin{figure}
\begin{center} \hspace{0.3cm}
	\begin{subfigure}[c]{0.4\textwidth}
		\centering
	\includegraphics[width=\textwidth]{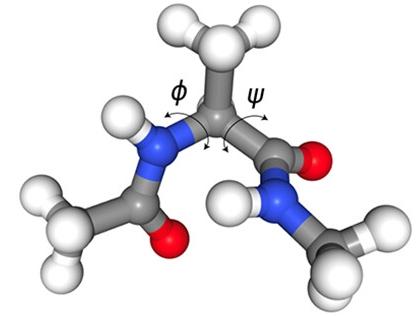} \vspace{-0.5cm}
		% \caption{}
	\end{subfigure} \hspace{0.5cm} 
	\begin{subfigure}[c]{0.4\textwidth}
		\centering 
	\includegraphics[width=\textwidth]{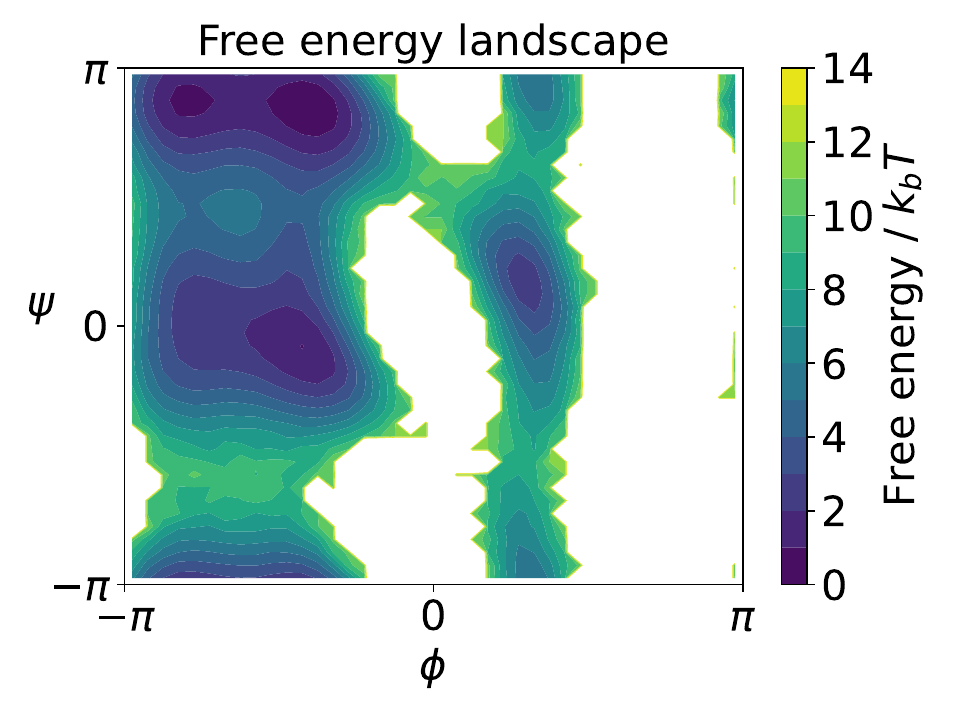}
		% \caption{}
	\end{subfigure} %\hspace{0.5cm} 
\caption{Graphical representation of the alanine dipeptide molecule on the left, and the reference free energy profile in two-dimensional dihedral angle space on the right.}
\label{fig:ala2-rep}
\end{center}
\end{figure}

Alanine dipeptide is a model system widely used in method development for simulation studies of macro-molecules. Figure \ref{fig:ala2-rep} shows the graphical representation of Alanine dipeptide. It is well known that the dynamical behavior of the molecule can be expressed in terms of the backbone dihedral angles $\phi$ and $\psi$, which constitute the two-dimensional reaction coordinate space defining the CG map $\xi$:
\begin{equation}
    \xi(x) = \begin{bmatrix}
        \phi(x) & \psi(x)
    \end{bmatrix}.
\end{equation}
We generated a $\qty{500}{\nano\second}$ simulation of the system in explicit water, the details of the simulation settings are summarized in Table~\ref{tab:md_setup} in the appendix.\\
The familiar free energy landscape of the system with respect to these two angles is shown in Figure~\ref{fig:ala2-rep},  displaying four minima, two on the left side, and two in the central part.

We apply the gEDMD algorithm with random Fourier features to find the meta-stable sets, and then use Algorithm~\ref{alg:cap} to learn the effective force and a state-dependent effective diffusion field in the dihedral angle space. Because of the periodicity of the CG coordinates, $\phi$ and $\psi$, the spectral measure corresponds to a periodic Gaussian kernel. Similar to the previous example, we tune the bandwidth of the kernel function using the VAMP-score.

\subsubsection{Meta-stability analysis}
\begin{figure}[ht]
\begin{center} \hspace{0.3cm}
	\begin{subfigure}[c]{0.4\textwidth}
		\centering
    \includegraphics[width=\textwidth]{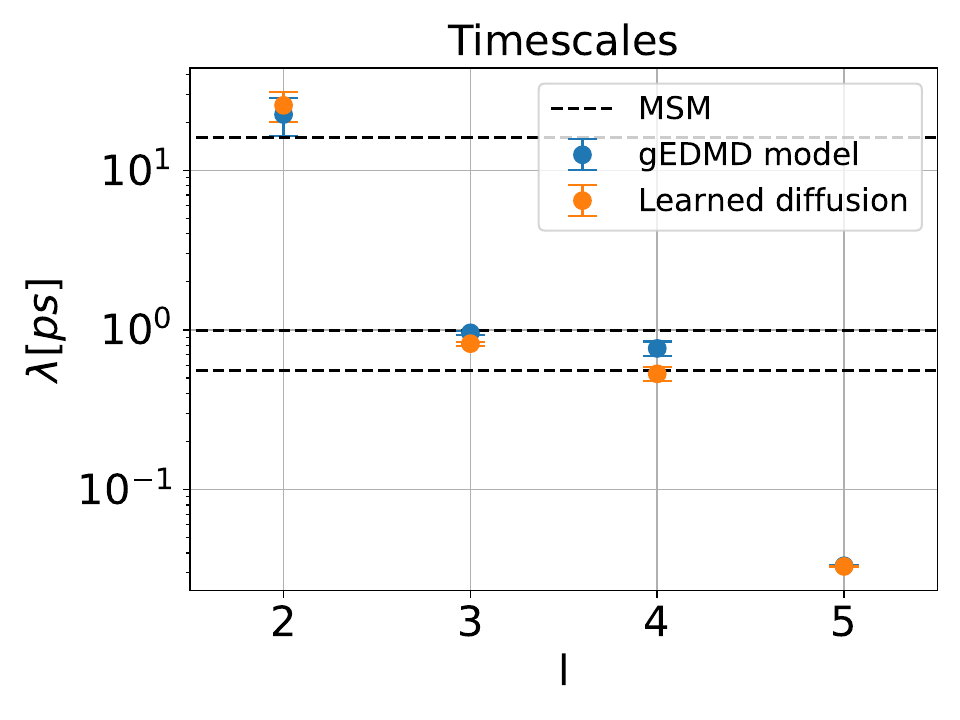}
		% \caption{}
	\end{subfigure} \hspace{0.3cm}
	\begin{subfigure}[c]{0.4\textwidth}
		\centering	\includegraphics[width=\textwidth]{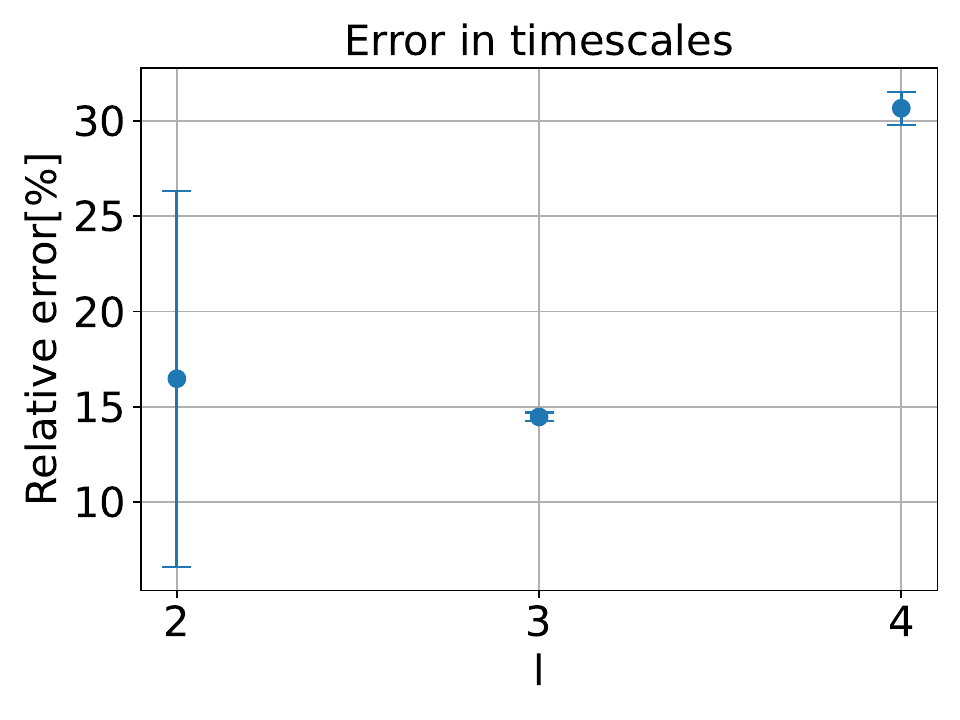}
% \caption{}
    \end{subfigure}
\caption{Approximation of generator for alanine dipeptide. The dominant timescales corresponding to the reference generator $\hat{\BL}_r$ and the learned generator $\hat{\BL}_\alpha^\xi$ built upon the learned effective diffusion on the left, and the relative error of these timescales compared to the reference is shown on the right.}
\label{fig:ala2-eigs}
\end{center}
\end{figure}

Figure~\ref{fig:ala2-eigs}a shows the leading finite timescales by taking reciprocals of the first three nonzero eigenvalues of the generator obtained from the gEDMD matrix $\hat{\BL}_r$ (error bars in the figure are generated by analyzing $5$ independent subsampled sets of the original data set, each comprising $50000$ samples).
The figure indicates the three dominant timescales which are corresponding to the four minima in the free energy landscape followed by a gap. In addition, we also show the timescales corresponding to the generator $\BL^\xi_\alpha$ based on the optimal effective diffusion, which agree well with the reference. As another way for validation of the timescales, we also computed the leading implied timescales using an MSM trained on the original simulation data, which does not rely on the overdamped assumption. These timescales exceed those of the generator model $\BL^\xi_\alpha$ by a uniform factor of about $100$, their re-scaled values are shown as black dashed lines. Thus, we observe that the dynamics in CG space based on the overdamped assumption is accelerated by a factor $100$ for this example, which confirms similar observations from previous studies. In principle, the exact transition timescales of the full simulation can be recovered by re-scaling the friction term of the CG model accordingly.

\subsubsection{Analysis of the CG dynamics}
\begin{figure}[h]
\begin{center} 
	\begin{subfigure}[c]{0.4\textwidth}
	\centering
	\includegraphics[width=\textwidth]{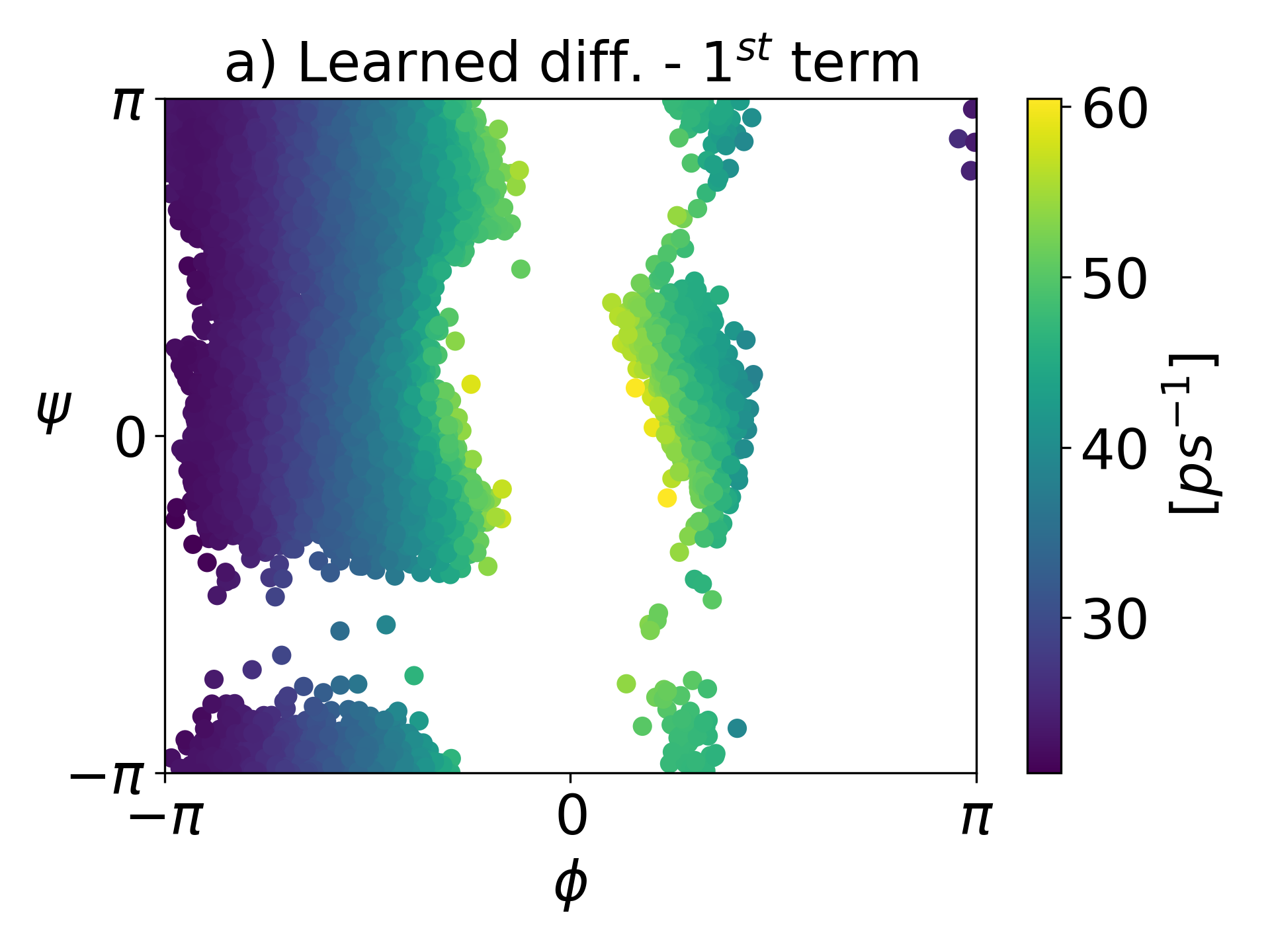}
	% \caption{}
    \label{fig:a-app1}
	\end{subfigure} %\hspace{0.5cm}
	\begin{subfigure}[c]{0.4\textwidth}
	\centering
	\includegraphics[width=\textwidth]{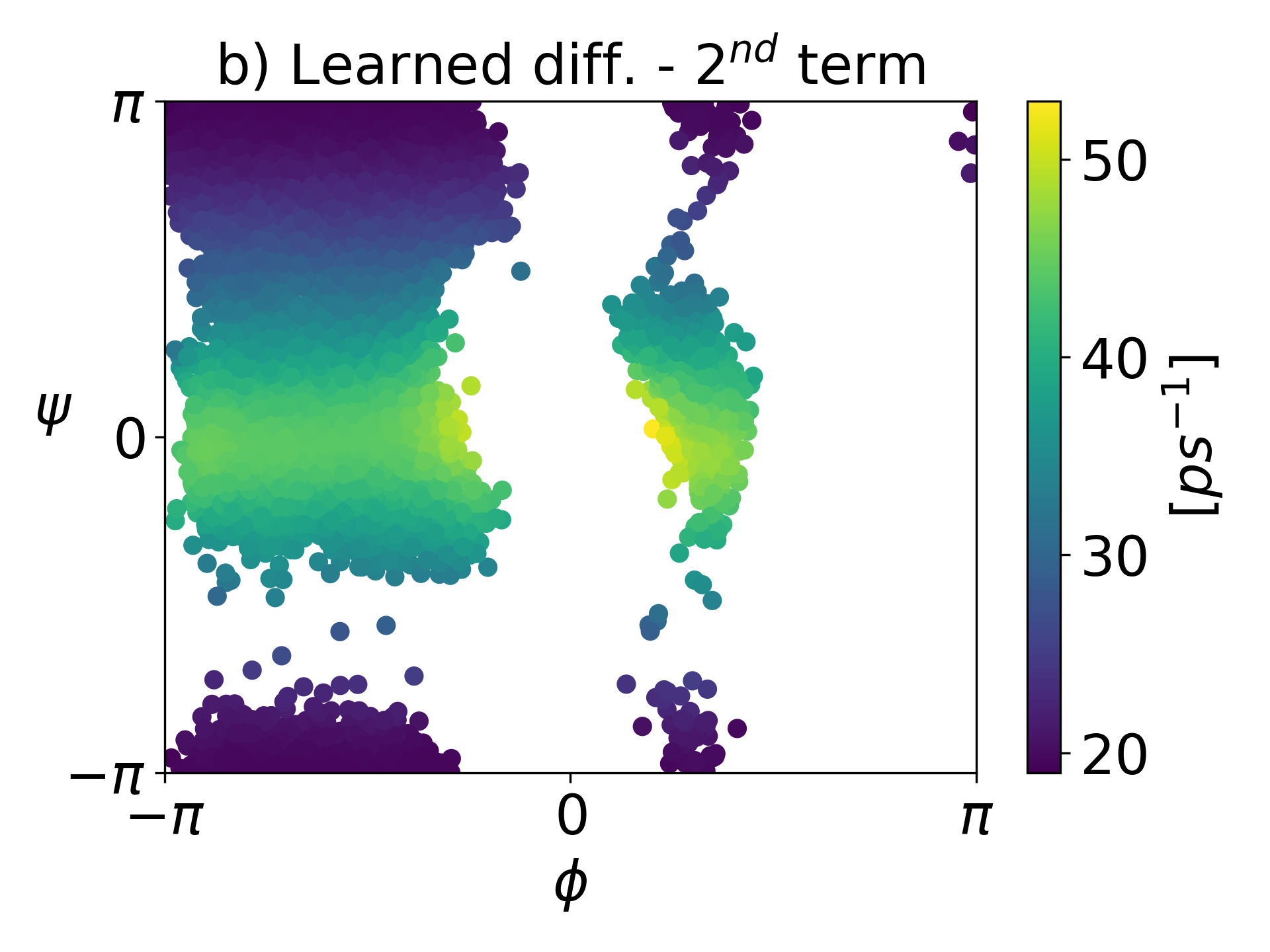}
	% \caption{}
    \label{fig:a-app2}
	\end{subfigure} %\hspace{0.5cm}
 	\begin{subfigure}[c]{0.4\textwidth}
	\centering
	\includegraphics[width=\textwidth]{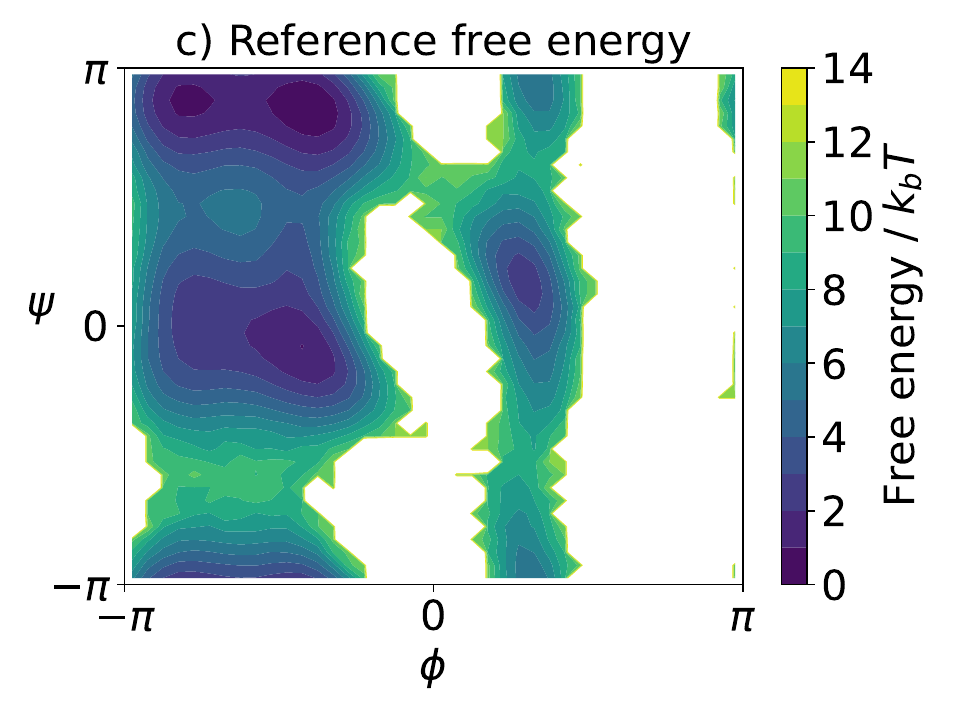}
	% \caption{}
 	\label{fig:pot-ref}
	\end{subfigure} %\hspace{0.5cm}v
	\begin{subfigure}[c]{0.4\textwidth}
		\centering
        \includegraphics[width=\textwidth]{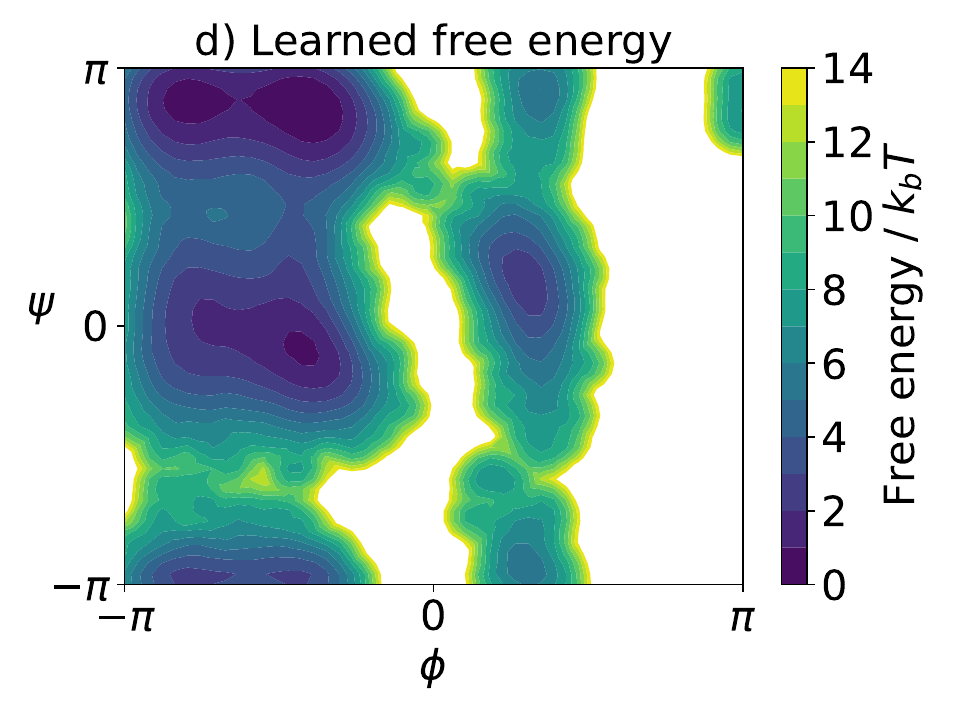}
	% \caption{}
 	\label{fig:pot-app}
	\end{subfigure}
\caption{The first (a) and second (b) diagonal terms of the learned diffusion covariance matrix $a$, the reference free energy surface (c) and the free energy surface learned via KDE (d).}
\label{fig:ala2-learned}
\end{center}
\end{figure}

\begin{figure}[ht] 
\begin{center}
    \begin{subfigure}[c]{0.4\textwidth} \vspace{0.2cm}
    	\centering
    	\includegraphics[width=\textwidth]{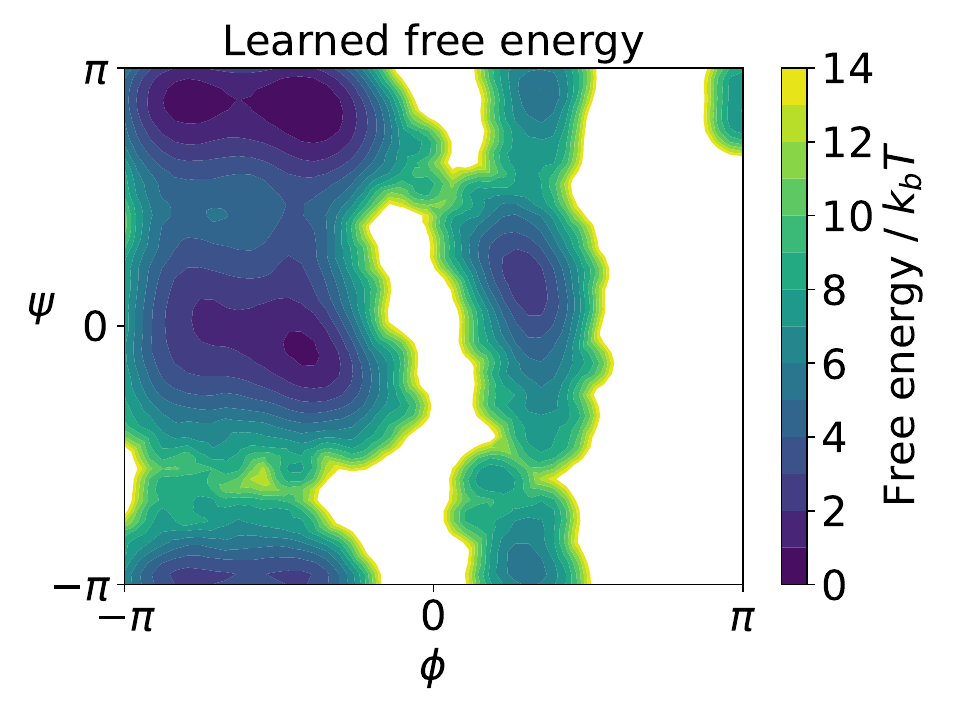}
         	% \caption{}
    	\label{fig:ala2-prop-pot}
    \end{subfigure}
    \begin{subfigure}[c]{0.4\textwidth} \vspace{0.2cm}
    	\centering
        \includegraphics[width=\textwidth]{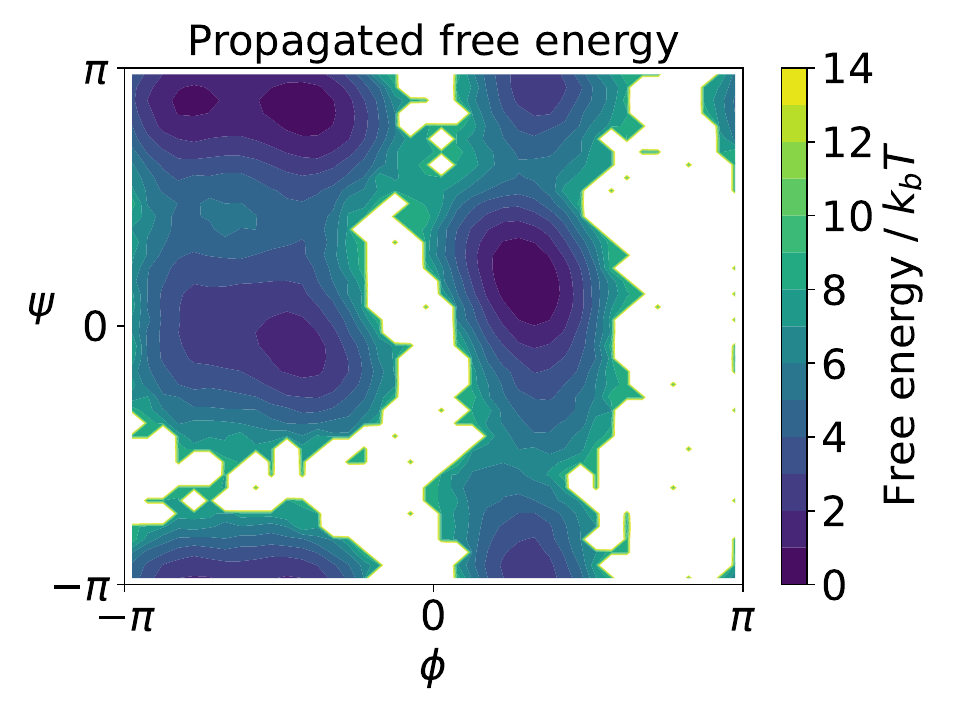}
    	% \includegraphics[width=\textwidth]{images/ala2_membership_prop.png}
    	% \caption{}
    	\label{fig:ala2-prop-mem}
    \end{subfigure}
    \caption{Left: Free energy surface learned via KDE. Right: estimated free surface from histogramming the simulated CG dynamics.}
    \label{fig:ala2-prop}
\end{center}
\end{figure}

\begin{figure}
    \begin{subfigure}[c]{0.45\textwidth} \vspace{0.2cm}
    	\centering
    	\includegraphics[width=\textwidth]{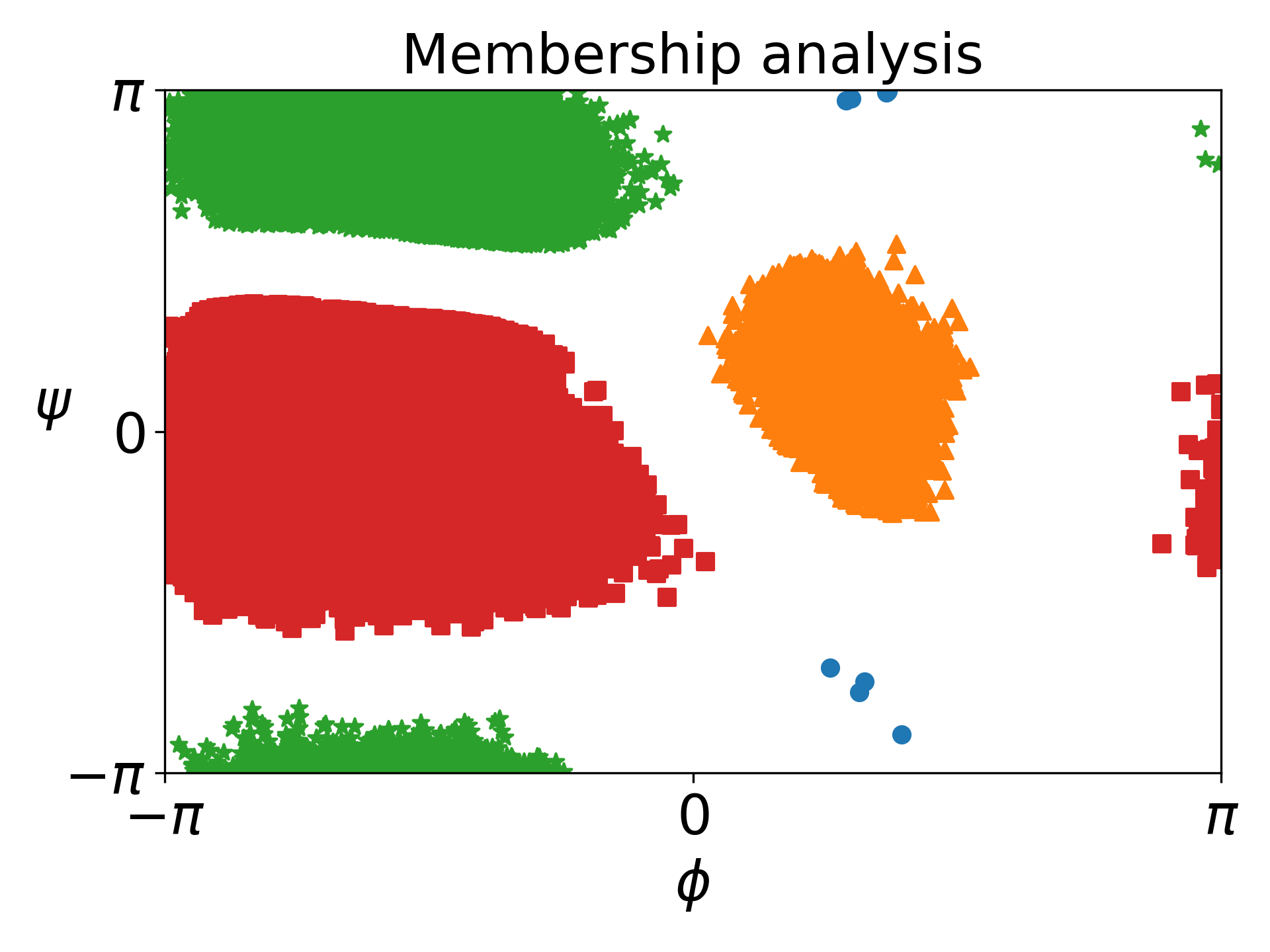}
    \end{subfigure}
    \begin{subfigure}[c]{0.45\textwidth} \vspace{0.2cm}
    	\centering
    	\includegraphics[width=\textwidth]{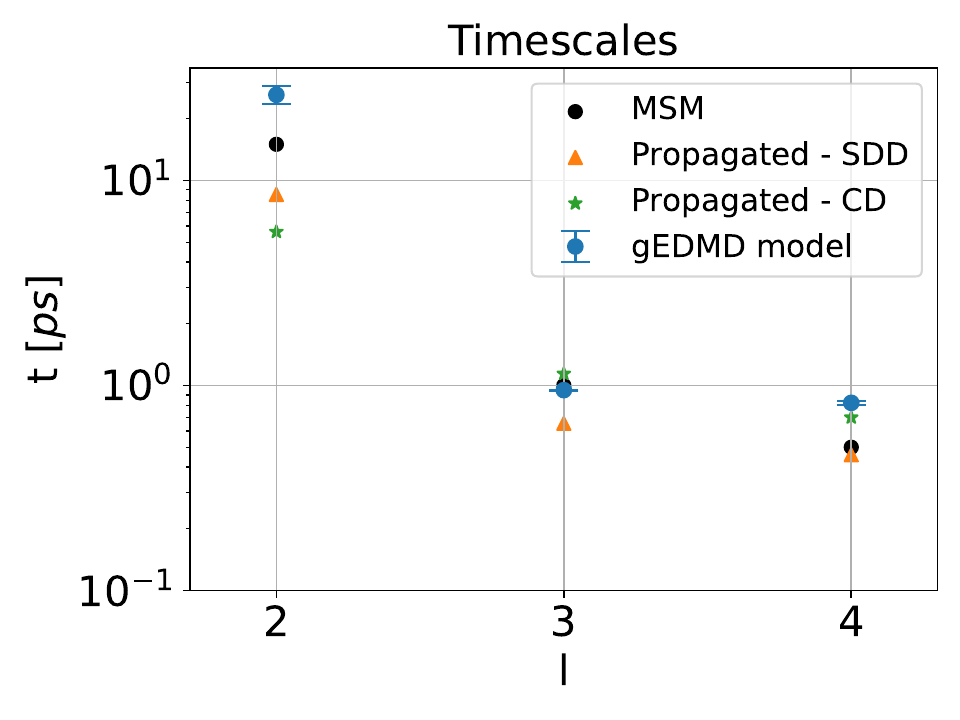}
    \end{subfigure}
    \caption{
    Kinetic consistency of the CG dynamics for alanine dipeptide. Left: PCCA+ membership analysis applied to simulation data of the CG dynamics. Right: slowest finite timescales calculated using an approximation of the generator from the reference dataset (blue) and the propagated CG dynamics with state-dependent diffusion (SDD, orange) as well as constant diffusion (CD, green), compared to those obtained via a Markov state model (black).} 
    \label{fig:ala2-prop-eigs}
\end{figure}

For this 2-dimensional coarse graining, we can express the diffusion field as a $2\times 2$ full matrix. For simplicity, however, we assume that the learned diffusion is a diagonal matrix. Figure~\ref{fig:ala2-learned} shows the first and second diagonal terms of the learned diffusion field based on $50000$ samples of the available dataset. 
To learn the effective potential, we found that the KDE method works best. The reference and learned effective free energy surfaces are depicted in Figures~\ref{fig:ala2-learned}c and ~\ref{fig:ala2-learned}d, respectively. It it noticeable that the learned free energy surface correctly captures all energetic minima and barriers up to some minor spurious behaviour close to the transition regions. We emphasize once again that this approximation could probably be improved further by using a more accurate learning method.\\
From the effective force and diffusion, one can compute the effective drift from which the SDE governing the dynamics in the CG space can be formed. We integrate the learned SDE for a short span of time ($\qty{100}{\pico\second}$). Figure~\ref{fig:ala2-prop}b shows the estimated free energy surface obtained from a histogram of the propagated dataset which is somewhat less accurate than the learned potential, most likely due to the short simulation time. Since we are mainly interested in kinetic properties, we estimate a new gEDMD model on the propagated dataset for the CG dynamics. We find that the four meta-stable states are correctly reproduced by a PCCA+ analysis of the propagated coarse grained SDE, as shown in the left panel of Figure~\ref{fig:ala2-prop-eigs}. In addition, we show the resulting transition timescales on the right of Figure~\ref{fig:ala2-prop-eigs}, compared to the ones corresponding to the learned generator built upon the original dataset, as well as the re-scaled MSM timescales. The results confirm that the two-dimensional CG dynamics with learned effective diffusion accurately recover the meta-stable states and transition timescales of the original dynamics, while adequately recovering their thermodynamic properties.\\

As a final analysis, we also generate a trajectory of the coarse grained SDE, but with the diffusion set to a constant. We choose the value of constant diffusion according to the average of the learned diffusion on the original dataset, resulting in $a\approx \qty{5.5}{{\pico\second}^{-1}}$. We also estimate a gEDMD model for these dynamics, and report the transition timescales in Figure~\ref{fig:ala2-prop-eigs}. The result shows that for alanine dipeptide, both the variable and constant diffusion fields lead to almost the same timescales.

\subsection{Chignolin}
\label{subsec:chignolin}
\subsubsection{System introduction}
\begin{figure}[ht] 
\begin{center}
    \begin{subfigure}[c]{0.45\textwidth} \vspace{0.2cm}
    	\centering
        \includegraphics[width=0.5\textwidth]{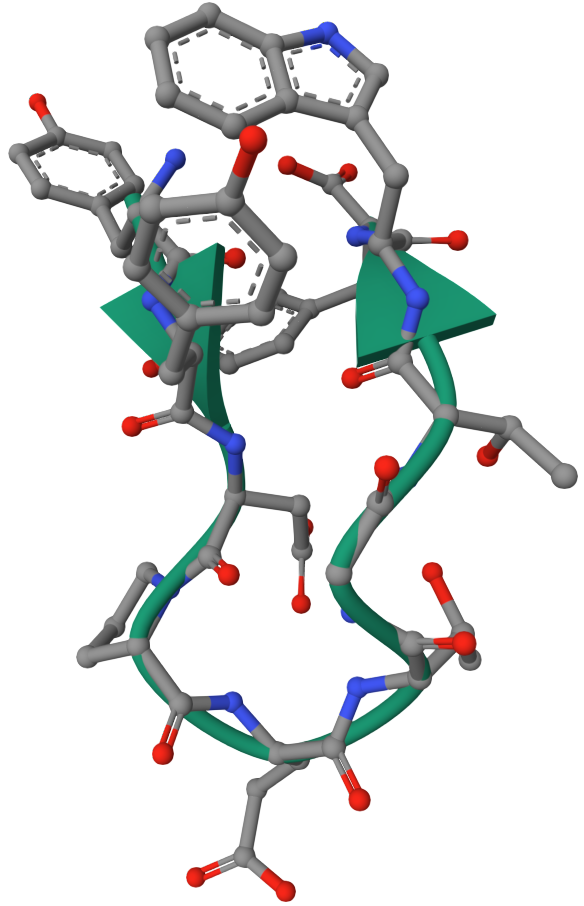}
    \end{subfigure}
    \begin{subfigure}[c]{0.45\textwidth} 
    	\centering	\includegraphics[width=\textwidth]{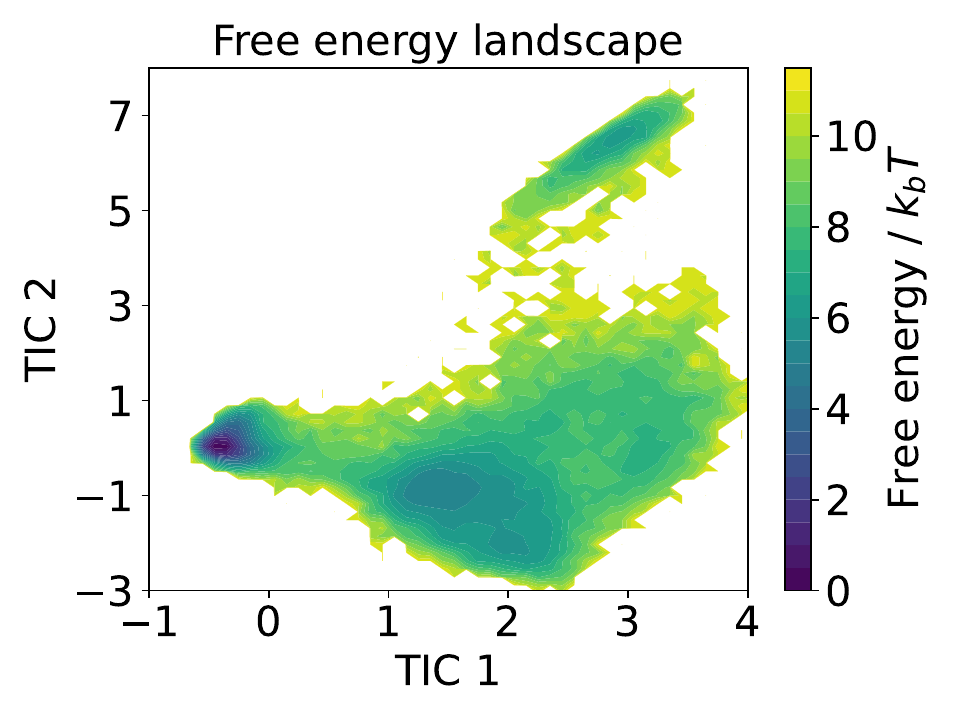}
    \end{subfigure}
    \caption[]{Graphical representation of CLN025 on the left \footnotemark, and the reference free energy surface in the two-dimensional TICA space on the right. The left-hand side minimum corresponds to the folded state, the bottom right minimum corresponds to the unfolded state and the top one associates to the misfolded state.}
    \label{fig:cln_gr}
\end{center}
\end{figure}

Finally, we apply the proposed method to the $"025"$ mutant of Chignolin (CLN025)~\cite{honda_crystal_2008} , which is a mini-protein consisting of $10$ amino acids. Figure~\ref{fig:cln_gr} shows the graphical representation of the molecule. The data for this example was obtained via simulation in $OpenMM$, see Ref.~\cite{charron_navigating_2023} for details of the setup. The dataset consists of $20$ independent trajectories each for $\qty{5}{\micro\second}$. In this example, we need to find a coarse graining function in a data-driven manner. To obtain the CG space, we performed TICA (Time-Lagged Independent Component Analysis) on a $45$-dimensional feature space comprising the $C^{\alpha}$ distances of the residues. As a result of TICA, we selected the first $2$ dominant components to constitute the RC space:
\begin{equation}
    \xi(x) = \begin{bmatrix}
        \mathrm{TIC}_1(x) & \mathrm{TIC}_2(x)
    \end{bmatrix}.
\end{equation}
By projecting the atomistic positional information of the system onto this $2$-dimensional TICA space and computing the histogram of the data, the free energy surface can be obtained, as shown in Figure~\ref{fig:cln_gr}.
The free energy surface shows three minima, representing the three conformational states of folded, unfolded and misfolded. We perform the same analysis as for the previous example.
% Similar to the previous example, we computed the local mean force and local diffusion according to Equations~\ref{eq:f_lmf} and~\ref{eq:loc-diff}, respectively.
\subsubsection{Meta-stability analysis}
\begin{figure}[ht]
\begin{center} \hspace{0.3cm}
	\begin{subfigure}[c]{0.45\textwidth}
		\centering
		\includegraphics[width=\textwidth]{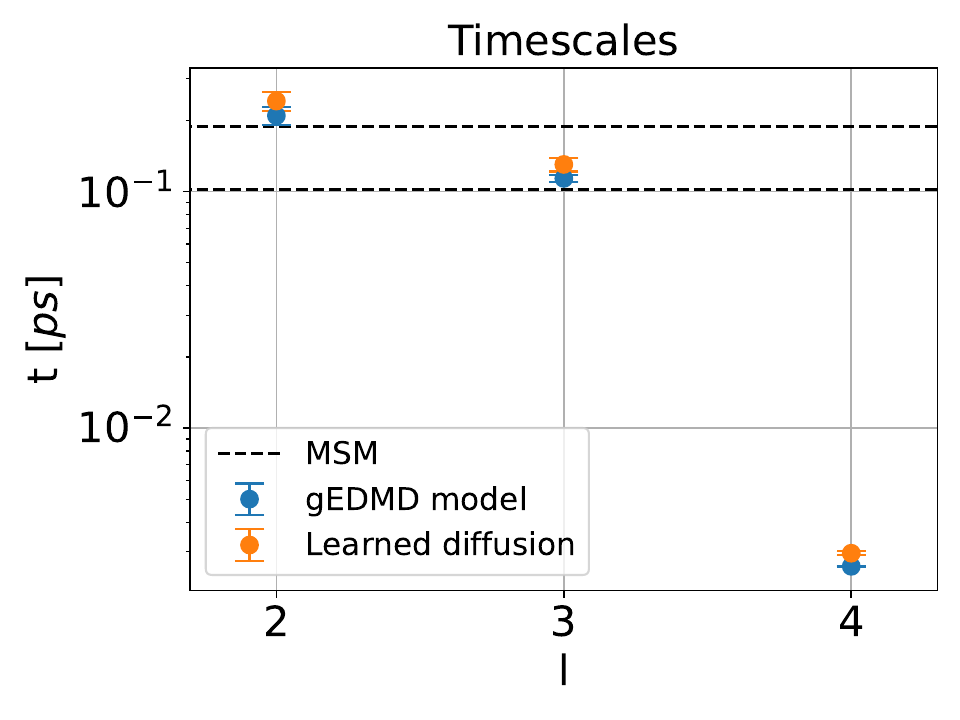}
		% \caption{}
	\end{subfigure} \hspace{0.3cm}
	\begin{subfigure}[c]{0.45\textwidth}
		\centering
        \includegraphics[width=\textwidth]{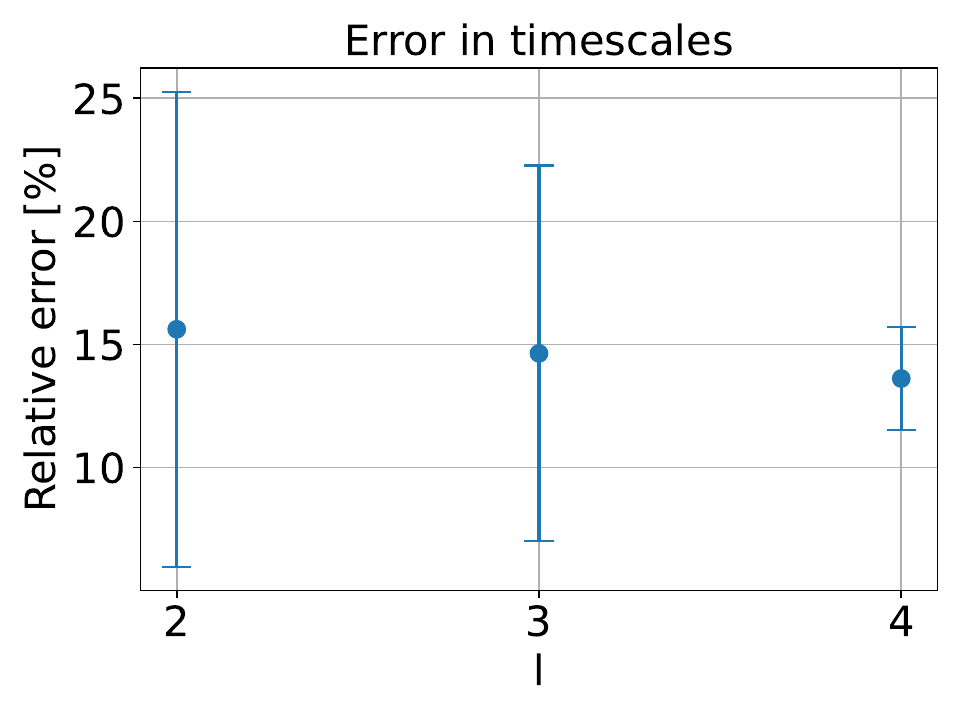}
% \caption{}
    \end{subfigure}
\caption{Approximation of generator for Chignolin. The slowest finite timescales corresponding to the reference generator $\hat{\BL}_r$ and the learned generator $\hat{\BL}_\alpha^\xi$ built upon the learned effective diffusion on the left, and the relative error on the right. The black dashed lines on the left indicate re-scaled timescales obtained from an MSM on the original simulation data.}
\label{fig:cln-eigs}
\end{center}
\end{figure}

To find the timescales of the system, we applied the gEDMD method with random Fourier features as before, and computed the eigenvalues of the generator model $\hat{\BL}_r$. Figure~\ref{fig:cln-eigs} shows the corresponding timescales of the system, which are the inverse of the generator's eigenvalues. The figure indicates the two leading timescales of the system (the first one associated to the stationary distribution is not shown) corresponding to the three meta-stable sets, followed by a spectral gap. Moreover, we show that the timescales of the CG generator $\BL^\xi_\alpha$ for the optimal effective diffusion are very similar,  the relative errors shown on the right of the same figure are sufficiently small. Also, we observe that the gEDMD timescales are once again uniformly re-scaled compared to the leading timescales of an MSM estimated on the original data, which is not subject to the overdamped assumption. The re-scaling factor is quite drastic this time, reducing micro-second timescales of the full system to less than pico-seconds for the CG dynamics. Nevertheless, as the re-scaling is again uniform, the original timescales can in principle be recovered by re-scaling the friction term. Error-bar figures were again generated by analyzing $5$ independent subsampled sets, each comprising $1.6\times10^5$ samples.
\footnotetext{The image is generated using Protein Data Bank in Europe platform.}
% Similar to the previous example, the CG dynamics is faster here as well, by a factor of $10^6$ that can be recovered by re-scaling of friction term. However, the ratio between the timescales is preserved.
\subsubsection{Analysis of the CG dynamics}
\begin{figure}[ht]
\begin{center} \hspace{0.3cm}
	\begin{subfigure}[c]{0.45\textwidth}
		\centering
	\includegraphics[width=\textwidth]{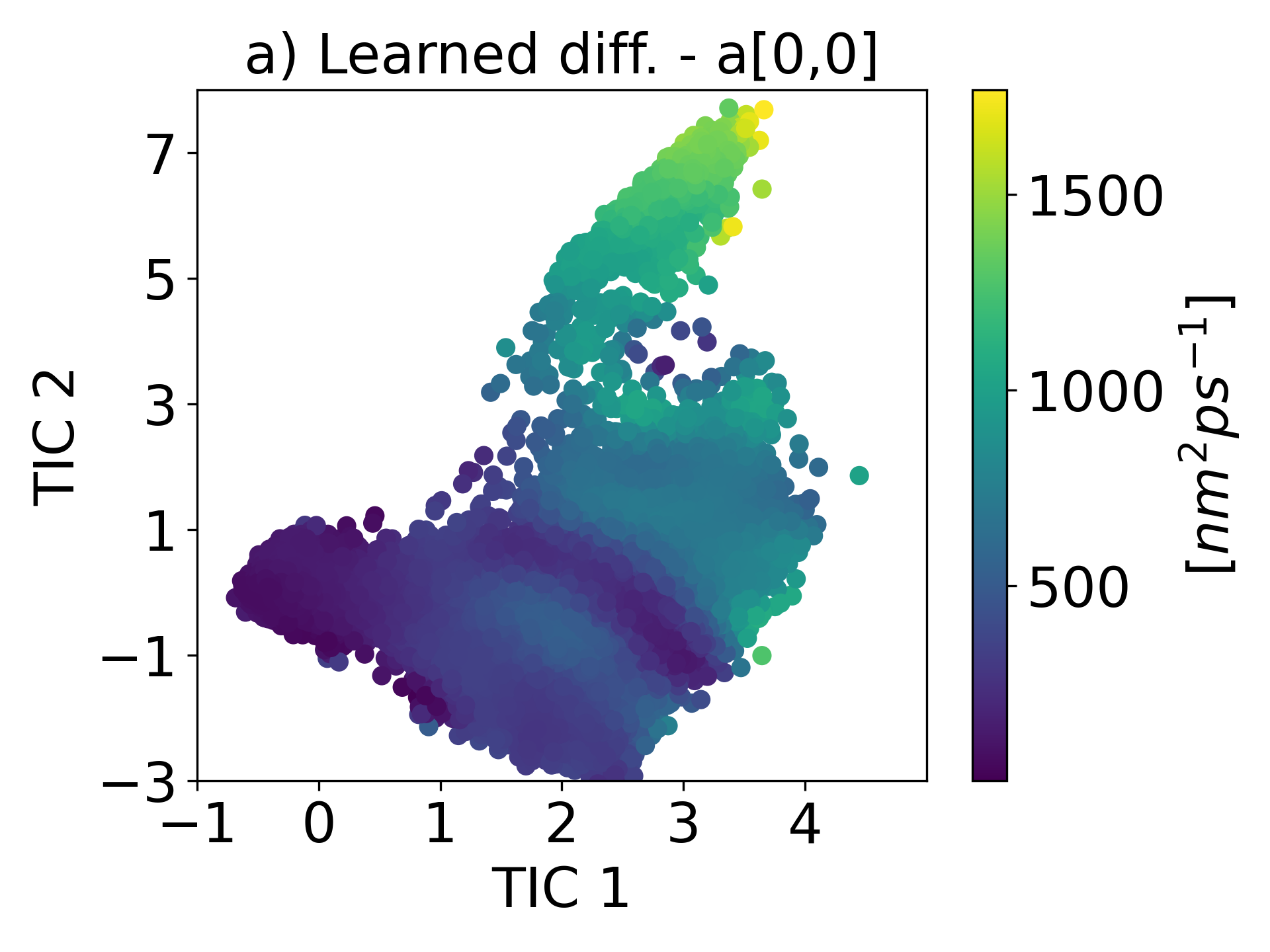}
		% \caption{}
	\end{subfigure} \hspace{0.3cm}
	\begin{subfigure}[c]{0.45\textwidth}
		\centering
	\includegraphics[width=\textwidth]{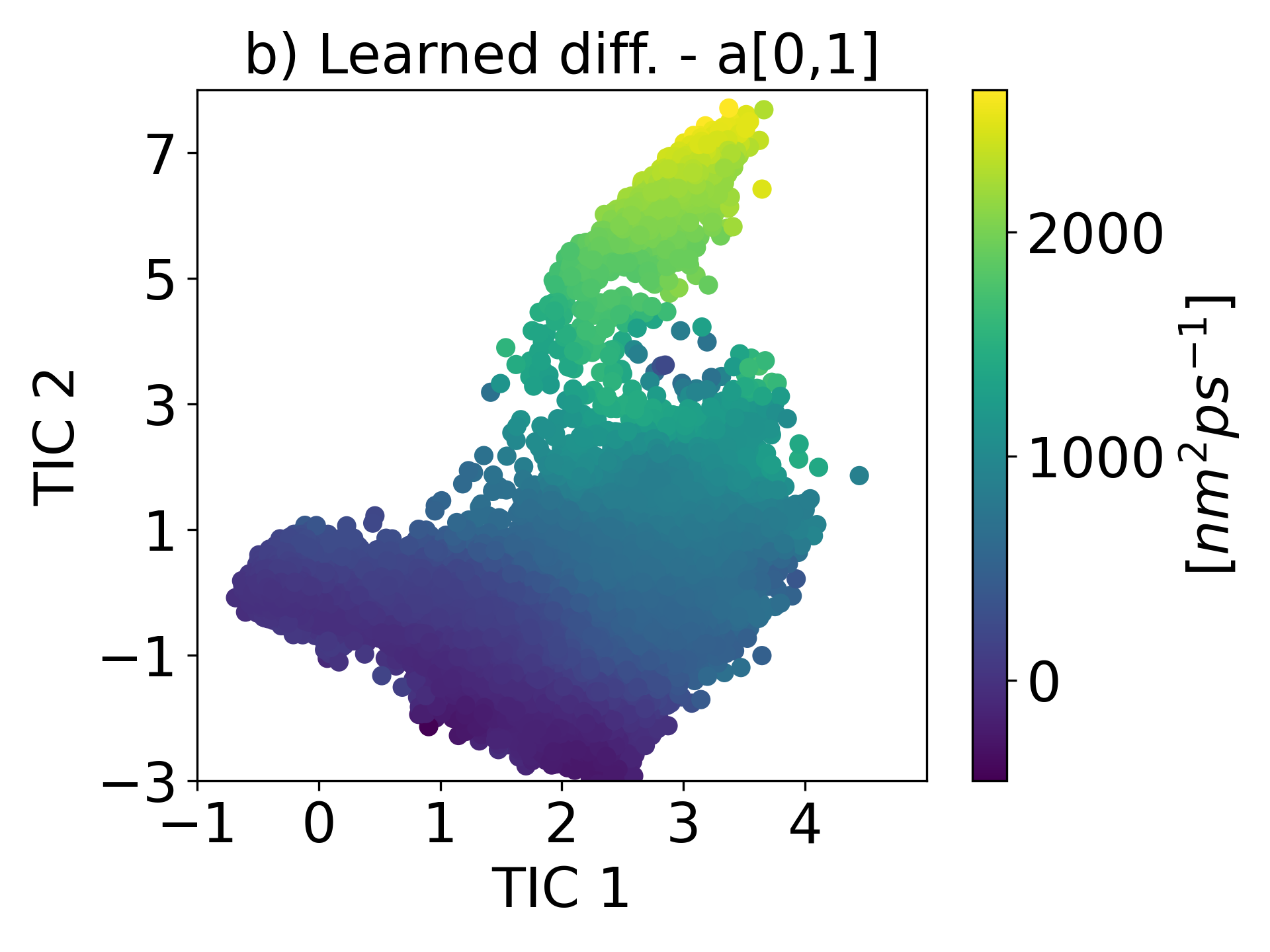}
% \caption{}
    \end{subfigure}
	\begin{subfigure}[c]{0.45\textwidth}
		\centering
	\includegraphics[width=\textwidth]{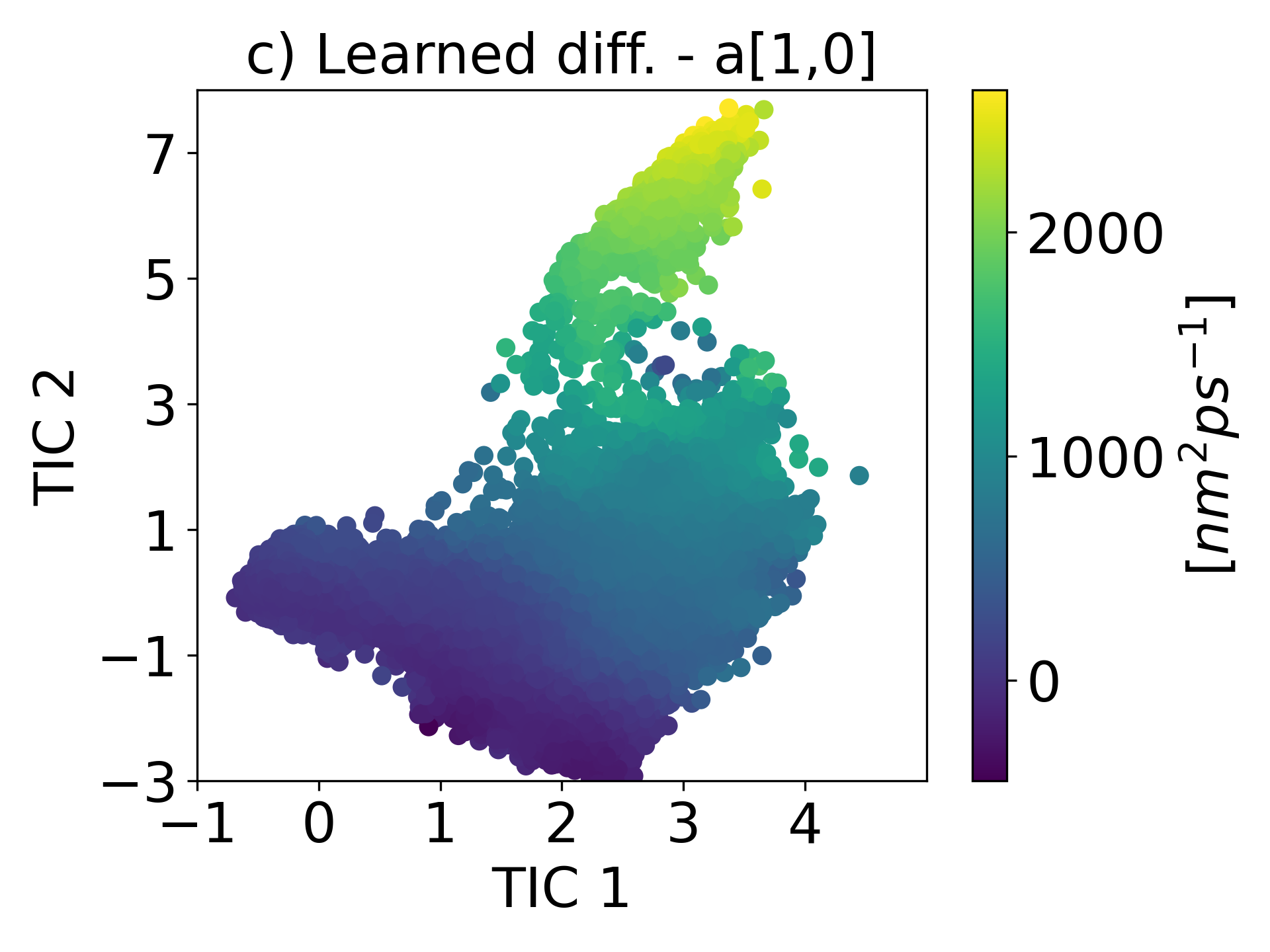}
		% \caption{}
	\end{subfigure} \hspace{0.3cm}
	\begin{subfigure}[c]{0.45\textwidth}
		\centering
	\includegraphics[width=\textwidth]{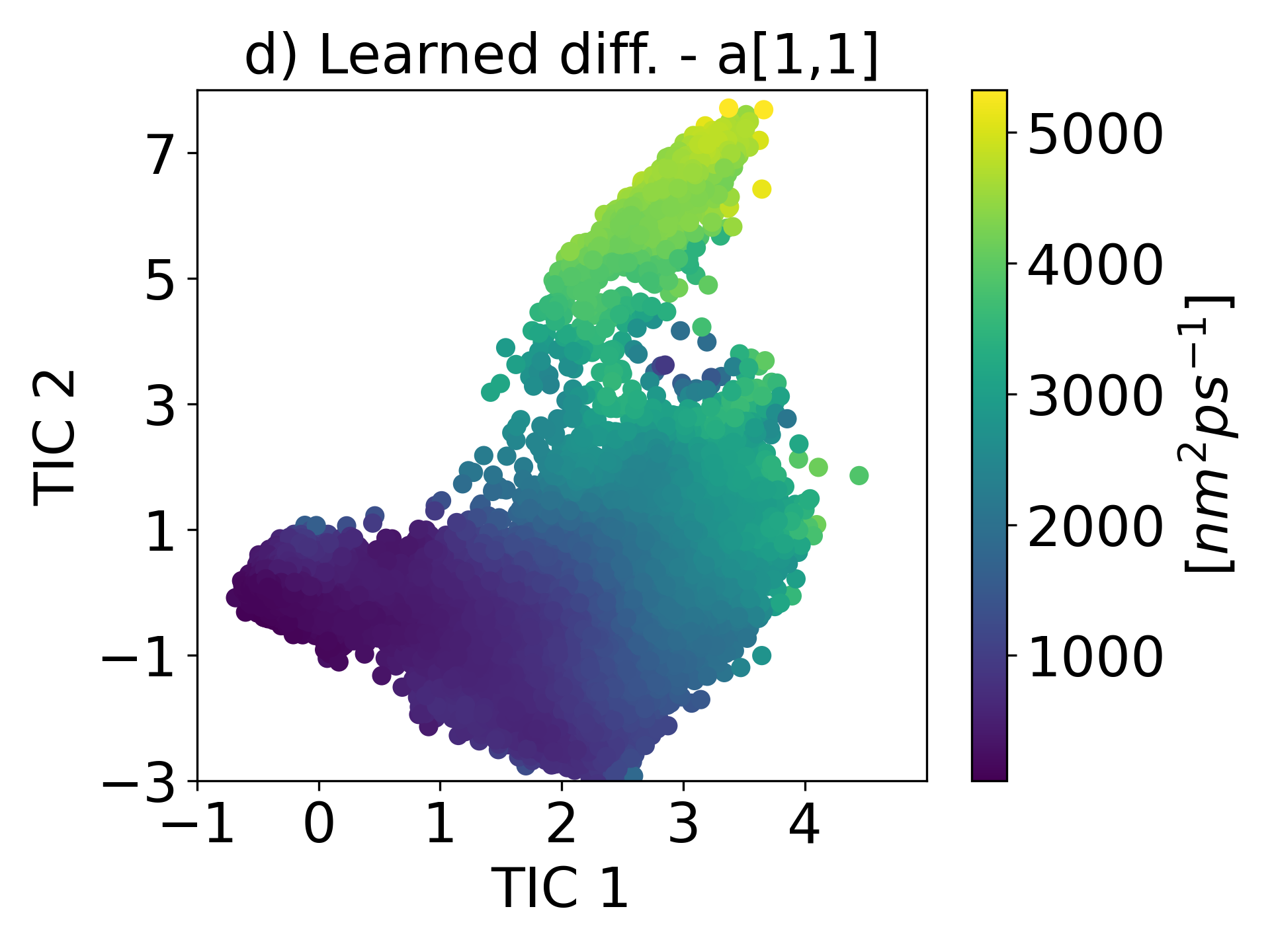}
% \caption{}
    \end{subfigure}
\caption{Components of the learned diffusion covariance matrix $a^\xi$ for Chignolin in its two-dimensional TICA space (note that the off-diagonal elements are symmetric).}
\label{fig:cln-learned-diff}
\end{center}
\end{figure}

\begin{figure}[ht]
    \begin{subfigure}[c]{0.45\textwidth}
        \centering
        \includegraphics[width=\textwidth]{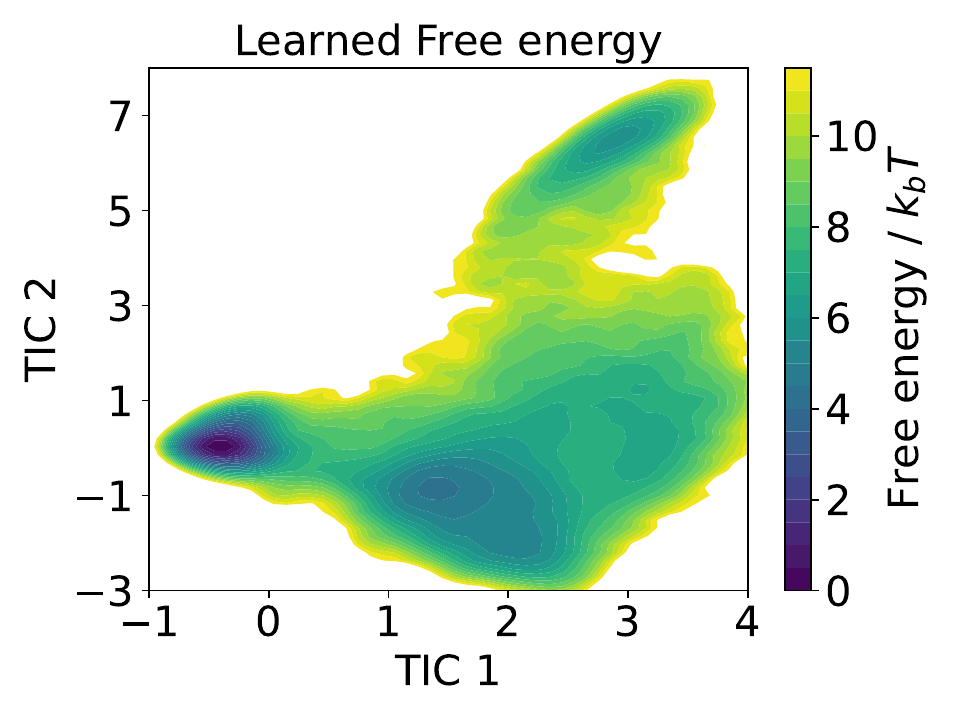} 
    \end{subfigure}
    \begin{subfigure}[c]{0.45\textwidth}
		\centering
	   \includegraphics[width=\textwidth]{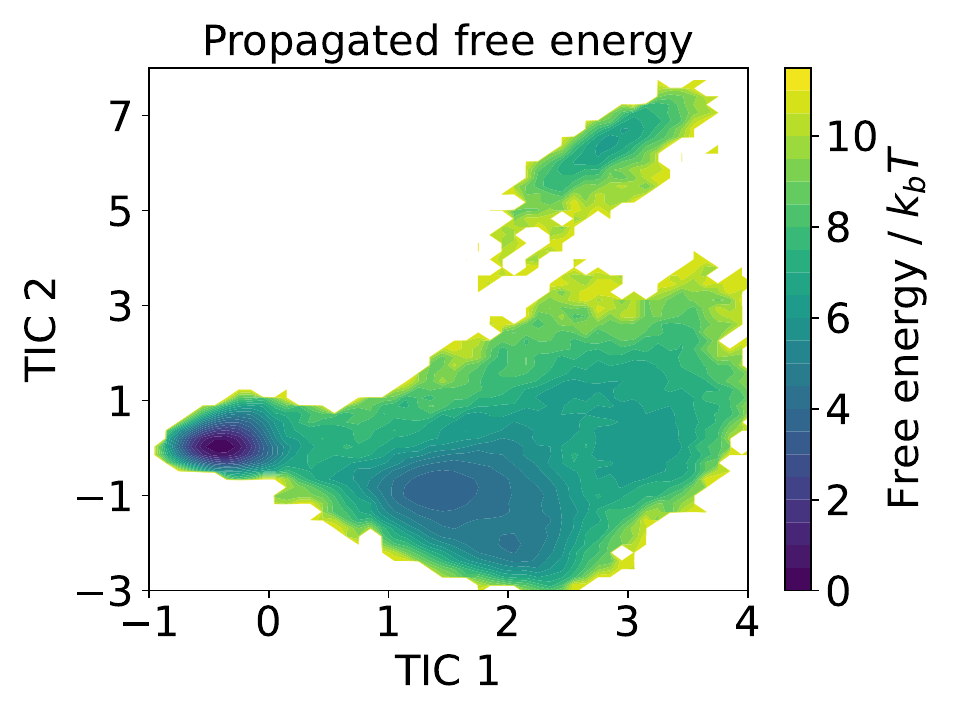}
		% \caption{}
	\end{subfigure}
\caption{Free energy surface in the two-dimensional TICA space for Chignolin, as learned by the KDE estimator on the left, and obtained from a histogram of the CG dynamics on the right.}
\label{fig:cln-learned-fes}
\end{figure}

Following the same procedure as in the previous examples, we learned a $2\times 2$ diffusion matrix in the CG space, but this time, we tested out a full non-diagonal diffusion field. Figure~\ref{fig:cln-learned-diff} shows the four elements of the learned diffusion matrix. In addition, the left panel of Figure~\ref{fig:cln-learned-fes} depicts the free energy surface learned by the KDE method, which is in satisfactory agreement with the reference one in Figure~\ref{fig:cln_gr}.

From the effective diffusion and potential energy, we compute the effective drift according to Equation~\ref{eq:drift_app}. We integrate the learned SDE for $\qty{10}{\pico\second}$ with a very small time step $\mathrm{dt} = 2\times10^{-5}$ ps. The right panel in Figure~\ref{fig:cln-learned-fes} shows the estimated free energy surface obtained from a histogram of the propagated CG dynamics. Once again, we find it in satisfactory agreement with the learned and the reference free energy in the CG space. Its accuracy could likely be improved by applying a more accurate learning method.

\begin{figure}
    \begin{subfigure}[c]{0.45\textwidth}
		\centering
	\includegraphics[width=\textwidth]{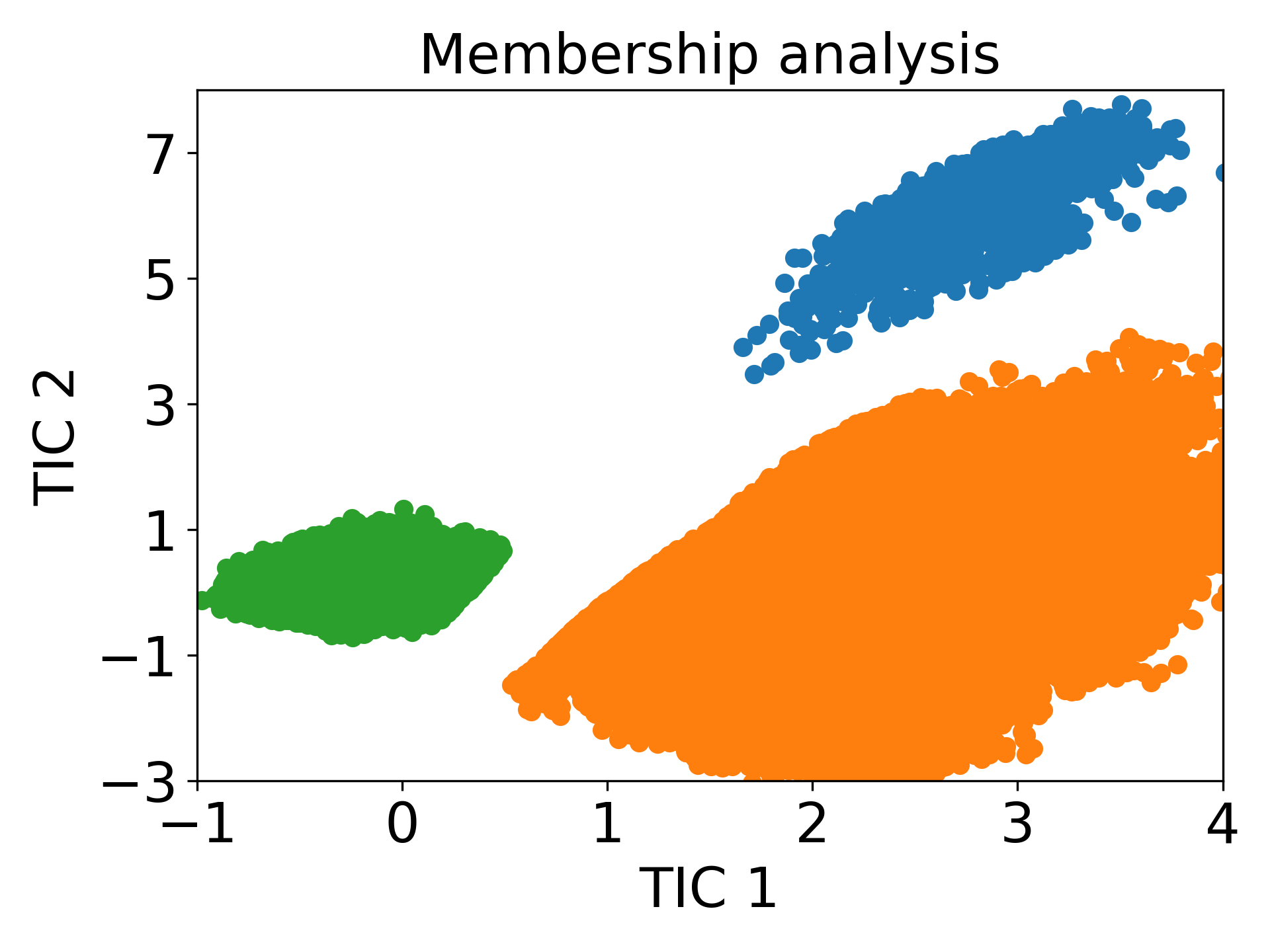}
% \caption{}
    \end{subfigure}
    \begin{subfigure}[c]{0.45\textwidth}
		\centering
        \includegraphics[width=\textwidth]{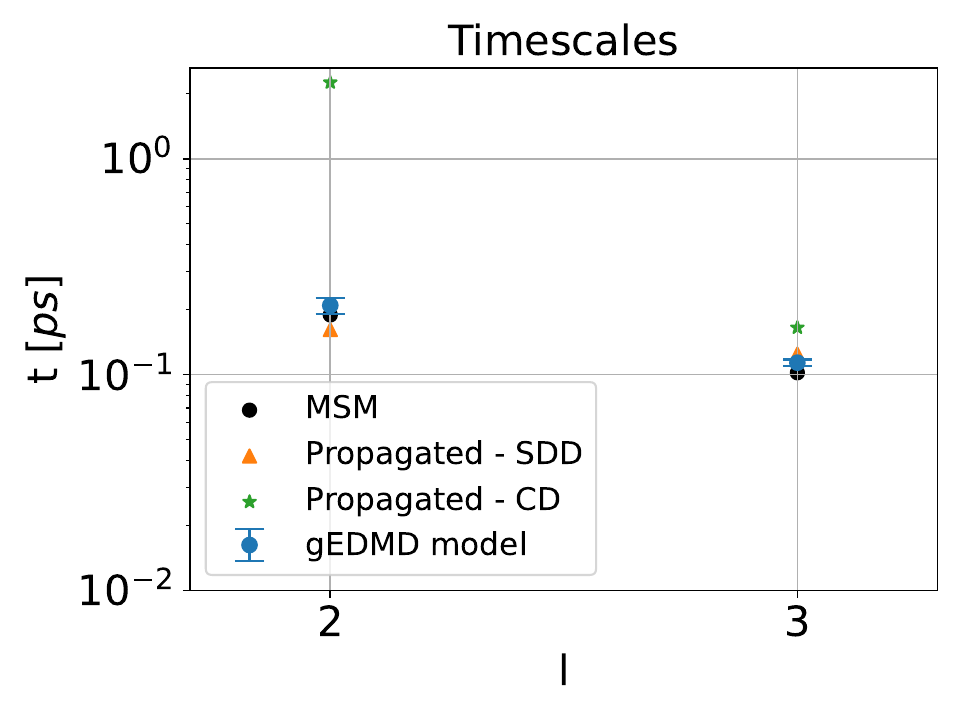}
    \end{subfigure}
    \caption{Kinetic consistency of the learned CG model for Chignolin. Left: PCCA+ states obtained from simulating the learned CG model. Right: slowest finite timescales of the system calculated using an approximation of the generator from the reference dataset (blue) and from the propagated CG dynamics (state-dependent diffusion in orange, constant diffusion in green). We also compare to re-scaled timescales from a Markov state model on the original simulation data (black).} 
    \label{fig:cln-prop-its}
\end{figure}

As we are mainly interested in kinetic properties, we compute a new gEDMD model on the propagated CG dynamics, and re-compute the associated eigenvalues and eigenvectors. The result of a PCCA+-analysis indicates that the correct meta-stable sets are recovered, as shown in the left panel in Figure~\ref{fig:cln-prop-its}. Likewise, the leading implied timescales estimated from the simulated CG dynamics are in good agreement with those of the original gEDMD model $\hat{\BL}_r$ and the re-scaled MSM timescales, both estimated from the original simulation data, as shown in the right panel of Figure~\ref{fig:cln-prop-its}.

Similar to the previous example, we also generate a separate trajectory based on a constant diffusion according to the average of the learned diffusion. We find that transition timescales for the constant diffusion are not well fitted to the reference. This is likely due to the fact that the variation of the diffusion field is stronger than in the previous example, and taking the mean leads to a slower system. This result confirms the need to learn a state-dependent diffusion field in the CG space to achieve kinetic consistency.

\section{Discussion}
We presented a novel approach to learn kinetically consistent coarse grained models for stochastic dynamics. We have introduced a learning method for the effective diffusion field in CG space, and shown how the kinetic properties of the CG dynamics can be evaluated by exploiting models for the Koopman generator (gEDMD algorithm). We have also shown that random Fourier features provide an efficient and flexible parametrization for both the effective diffusion and the gEDMD model. By means of three examples, a two-dimensional model potential and two datasets of molecular dynamics simulations, we showed that the effective dynamics in low-dimensional reaction coordinate spaces are able to reproduce both thermodynamic and kinetic quantities of the full dynamics accurately.\\
For the molecular examples, we have relied on the overdamped assumption to parametrize reversible CG dynamics. We have seen that this assumption leads to a uniform acceleration of the CG dynamics compared to the full system. The re-scaling factor can be estimated numerically by comparing the gEDMD model to a kinetic model that does not rely on the overdamped assumption. We used MSMs in this paper, but note that a more general EDMD model (e.g. using random features) would work just as well.\\
In this study, we used long equilibrium simulations to train CG models. However, one of the appealing aspects of the generator EDMD approach is that it only requires Boltzmann samples. As has been pointed out in previous studies, these samples can also be obtained from biased sampling simulations~\cite{lucke_tgedmd_2022,devergne_biased_2024} , or by employing generative models~\cite{moqvist_thermodynamic_2025} .\\
Among other topics, future work will focus on applying the formalism to higher-dimensional and more transferrable CG coordinates, for example C-alpha models. Another topic is the construction of CG models that can explicitly account for the underdamped structure of the full system, or that can incorporate memory terms, which were entirely disregarded in our study. Moreover, one can also try to simultaneously optimize the CG mapping $\xi$ along with the parameters of the CG model, for instance by balancing the VAMP score versus the complexity of the CG model.

\section*{Acknowledgments}
We thank the Theoretical and Computational Biophysics Group at Freie Universität Berlin for sharing the simulation data of the Chignolin mini-protein.

\section*{Data Availability Statement}
Codes and data to reproduce the results and figures shown in this manuscript are available from the following public repository: \url{10.5281/zenodo.13808938}.
\section{Appendix}
\label{sec:appen}
\subsection{Notation}
\label{subsec:not}
The most important notation used in the manuscript is summarized in Table~\ref{tab:not}.
\begin{table}[H]
  \begin{center}
    \caption{Overview of notation}
    \begin{tabular}{|c|c|}
      \hline
        $X_t$ &  stochastic process \\
        \hline
        % $\bX$ & state space\\
        % % \hline
        % $k,\Phi $ & kernel and associated feature map \\
        % \hline
        % $\bH$ & reproducing kernel Hilbert space \\
        \hline
        $\cK^{\tau}$  & Koopman operator with lag time $\tau$ \\
        \hline
        $\cL$ & generator of the Koopman operator \\
        \hline
        % $\hat{A}$ & empirical estimate of operator $A$ \\
        % \hline
        $\bh$ & reduced basis set from whitening transformation\\
        \hline
        $\hat{\BL}$, $\hat{\BL}_r$ & generator matrix and reduced generator matrix\\
        \hline
        $\sigma^\xi_\alpha$ & effective diffusion parameterized by $\alpha$\\
        \hline
        $\hat{\BL}^\xi_\alpha$ & effective generator matrix for diffusion with parameters $\alpha$ \\
        \hline
        $F$, $F^\xi$ & potential and effective potential \\
        \hline
        $f^\xi_{\rm lmf}$, $a^\xi_{\rm loc}$ & local mean force and local diffusion\\
        \hline
        % $\nabla_1 f$ & gradient of $f$ with respect to the first argument \\
        % \hline
        % $\nabla_2 f$ & gradient of $f$ with respect to the second argument \\
        % \hline
        $A\con{i}{j}B$ & contraction of dimensions $i$ and $j$ of arrays $A$ and $B$.\\
        \hline
      \end{tabular}
    \label{tab:not}
  \end{center}
\end{table}

\subsection{VAMP-score}
\label{subsec:vamp}
\begin{figure}[H]
    \begin{center} 
    \begin{subfigure}[c]{0.45\textwidth}  
    \includegraphics[width=\textwidth]{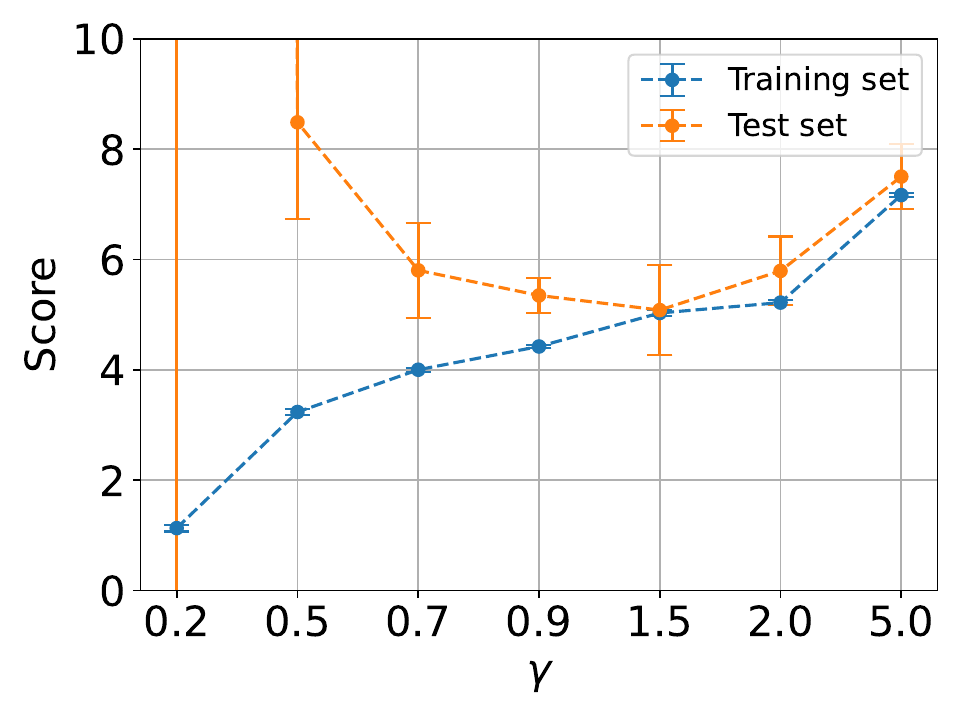}
    % \caption{}
    % \label{fig:lemon-pot}
    \end{subfigure}
    \begin{subfigure}[c]{0.45\textwidth}  
    \includegraphics[width=\textwidth]{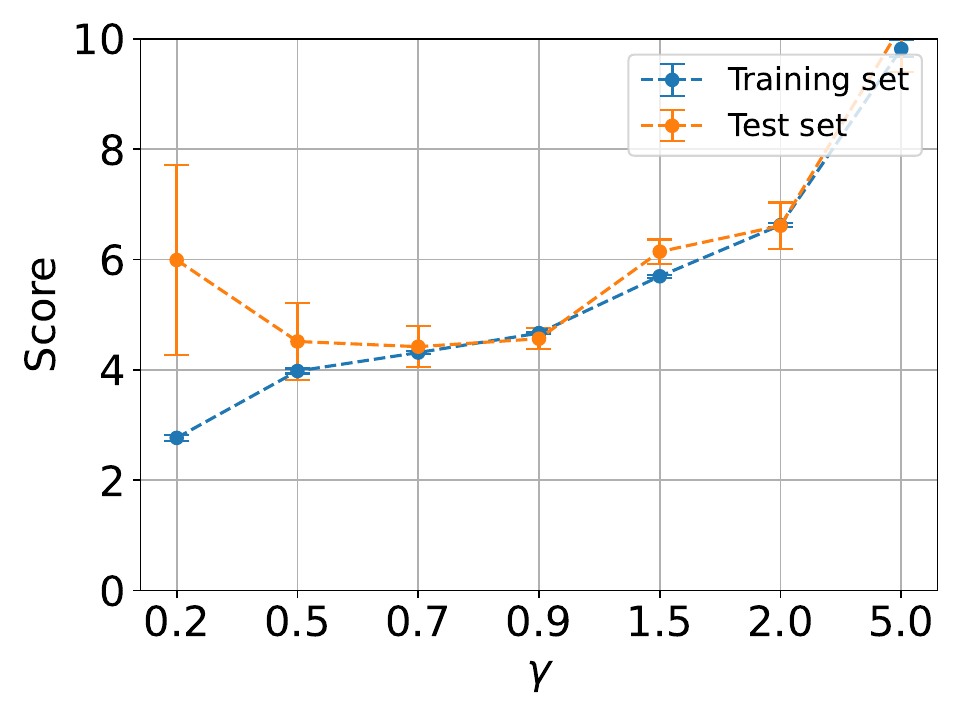}
    % \caption{}
    % \label{fig:lemon-member}
    \end{subfigure}
    \caption{VAMP-score analysis for the Lemon slice example using periodic Gaussian kernel on the left and Gaussian kernel on the right.}
    \label{fig:lemon-vamp}
    \end{center}
\end{figure}

\begin{figure}[H]
    \begin{center} 
    \begin{subfigure}[c]{0.45\textwidth}  
    \includegraphics[width=\textwidth]{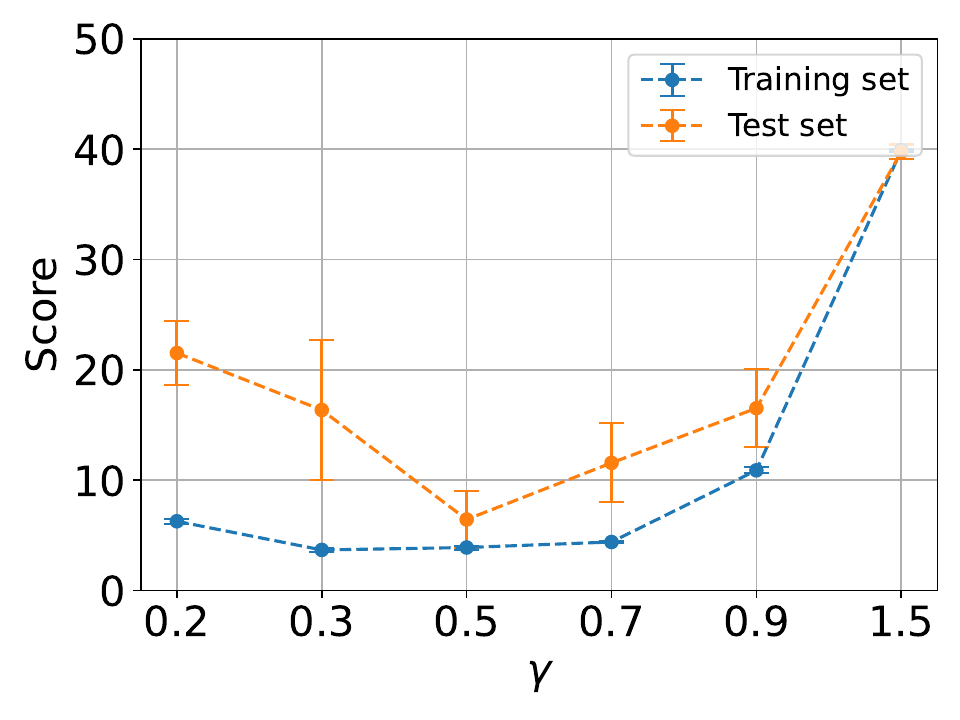}
    \end{subfigure}
    \begin{subfigure}[c]{0.45\textwidth}  
    \includegraphics[width=\textwidth]{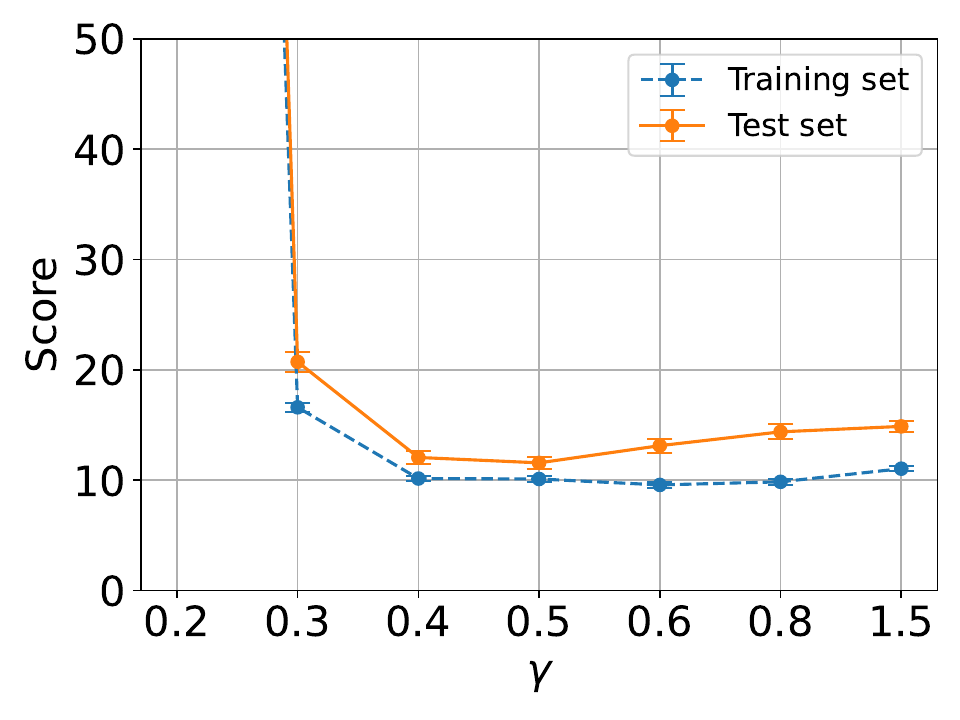}
    % \caption{}
    % \label{fig:lemon-member}
    \end{subfigure}
    \caption{VAMP-score analysis for alanine dipeptide on the left and Chignolin on the right.}
    \label{fig:ala2_cln_vamp}
    \end{center}
\end{figure}

We tune hyper-parameters of the proposed method based on \textit{VAMP variational principle} proposed in \cite{wu_variational_2020} , stating that for reversible systems, the $k$ dominant eigenvalues of the Koopman generator can be obtained by a minimization problem
\begin{equation}
    \sum_{i=1}^k \lambda_i = \min_{\phi_0,...,\phi_k} \sum_{i=1}^k \innerprod{\phi_i}{\cL \phi_i}_{\mu} 
\end{equation}
where the $\phi_i$ are orthogonal functions. We use this variational principle to optimize the kernel bandwidth giving rise to the spectral measure used in the context of our proposed method. To do this robustly and avoid overfitting, we make use of standard cross validation scheme by introducing $40\%$ of dataset as test set. Figure \ref{fig:lemon-vamp} shows the result for the lemon slice example, using Gaussian and periodic Gaussian kernels.
% The larger the bandwidth, the higher the scores for both training set and test set, as we are proposing a wide kernel function which fails to capture the difference between close samples. On the other side of the spectrum, the score decreases monotonically for the training set due to the fact that by decreasing the bandwidth, we are localizing the kernel functions on the training set, so we are improving the score for this set which makes the optimized kernel overfitting the data and loses the score for the test set. \\
We applied the same procedure for optimizing the bandwidth for alanine dipeptide and Chignolin. Figure \ref{fig:ala2_cln_vamp} shows the result for optimization of the bandwidth.
As the figure indicates, there exist a range of the bandwidth $\gamma$ that we can safely choose the bandwidth from.
% Here, due to approximation of the generator using RFF, lowering the bandwidth beyond a range increases the score even for the training set. In both examples, we are interested in taking the bandwidth corresponding to the smallest score, but having a range of acceptable values for the bandwidth.  

\subsection{Simulation Settings for Alanine Dipeptide}
For the example of alanine dipeptide, we used the Gromacs~\cite{bekker_gromacs-parallel_1993} simulation software to produce a $\qty{500}{\nano\second}$ simulation. The details of the input setting we used for running the simulation is summarized in Table~\ref{tab:md_setup}. 
\begin{table}[H]
  \begin{center}
    \caption{Experiment setup}
    \begin{tabular}{|c|c|}
    \hline
        Force Field & {AMBER99SB-ILDN}\\
      \hline
        Temperature & $\qty{300}{\kelvin}$ \\
        \hline
        Time constant ($1/\gamma$) & $\qty{0.2}{\pico\second}$ \\
        \hline
        Integrator & Langevin dynamics \\
        \hline
        Time step & $\qty{2}{\femto\second}$ \\
        \hline
        Simulation time  & $\qty{500}{\nano\second}$ \\
        \hline
        Export data frequency & $\qty{100}{\femto\second}$ \\
        \hline
      \end{tabular}
    \label{tab:md_setup}
  \end{center}
\end{table}

\pagebreak

% \bibliography{biblio}
\bibliography{refs_cg}

\pagebreak

\begin{figure}
    \begin{center} 
    \includegraphics[width=\textwidth]{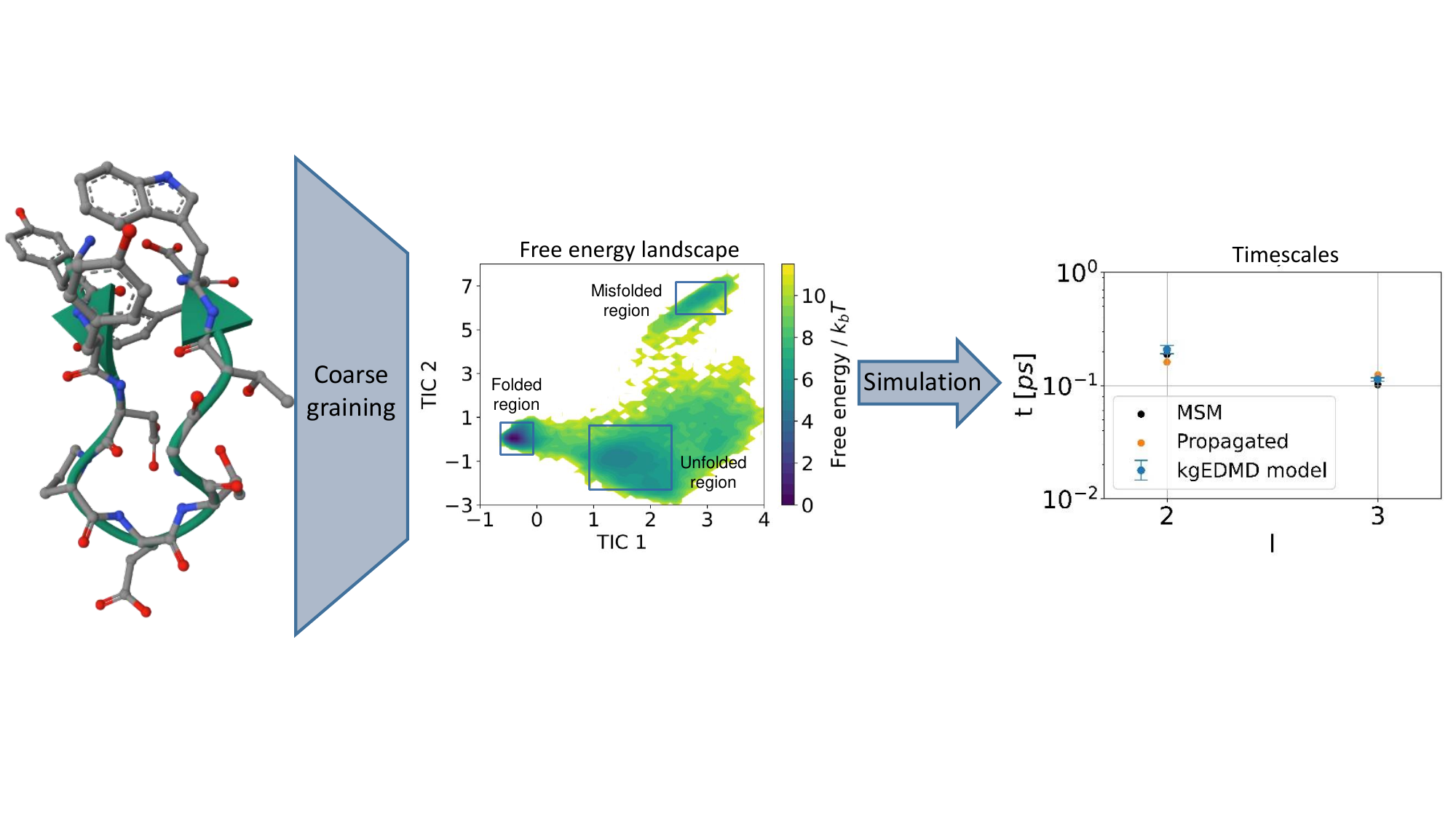}
    \caption{
    TOC Graphic} 
    \end{center}
\end{figure}

\end{document}